%% file: ms_cbla_fin.tex
\documentclass{emulateapj}

\slugcomment{Accepted version from February 21, 2020}
\shorttitle{Coronal Broad Lyman $\alpha$ Absorbers}
\shortauthors{Philipp Richter}

\begin{document}

\title{Hot gas in galaxy halos\\ traced by coronal broad Lyman $\alpha$ absorbers}

\author{Philipp Richter\altaffilmark{1,2}
}

\affil{$^1$Institut f\"ur Physik und Astronomie, Universit\"at Potsdam,
Haus 28, Karl-Liebknecht-Str.\,24/25, 14476 Golm (Potsdam),
Germany}
\affil{$^2$Visiting Erskine Fellow at the University of Canterbury, 
Department of Physics and Astronomy, Christchurch 8020, New Zealand}

%

\begin{abstract}

We explore the possibility to systematically study the extended, hot gaseous halos
of low-redshift galaxies with Coronal Broad Ly\,$\alpha$ Absorbers (CBLAs). 
These are weak, thermally broadenend H\,{\sc i} absorption lines arising from the 
tiny fraction of neutral hydrogen that resides in the collisionally ionized,
million-degree halo gas in these galaxies.
Using a semi-analytic approach, we model the spatial density and temperature distribution 
of hot coronal gas to predict strength, spectral shape, and cross section of CBLAs as a function 
of galaxy-halo mass and line-of-sight impact parameter. For virial halo masses in the range
log $(M/M_{\sun})=10.6-12.6$, the characteristic logarithmic CBLA H\,{\sc i} column densities
and Doppler parameters are log $N$(H\,{\sc i}$)=12.4-13.4$ and $b$(H\,{\sc i}$)=70-200$
km\,s$^{-1}$, indicating that CBLAs represent weak, shallow spectral features that
are difficult to detect. Yet, the expected number density of CBLAs per unit redshift 
in the above given mass range {\rm is $d{\cal N}/dz$(CBLA$)\approx 3$}, implying
that CBLAs have a substantial absorption cross-section.
We compare the model predictions with a combined set of
ultraviolet (UV) absorption-line spectra from HST/COS and HST/STIS that trace the halos 
of four low-redshift galaxies. We demonstrate that CBLAs already might have been detected in these 
spectra, but the complex multi-component structure and the limited 
signal-to-noise ratio (S/N) complicate the interpretation of these CBLA candidate systems. 
Our study suggests that CBLAs represent a very interesting 
absorber class that potentially will allow us to further explore the hot coronae of
galaxies with UV spectral data.

\end{abstract}

%

 
%

\section{Introduction}

Spiral galaxies like the Milky Way are believed to be surrounded
by large amounts of diffuse gas that is gravitationally bound
to a galaxy's potential well and extends to its virial radius (and beyond).
The presence of this so-called circumgalactic medium (CGM) can
be understood in the framework of $\Lambda$CDM galaxy formation models
(e.g., White \& Frenk 1991), which predict that diffuse gas in 
cosmological filaments is accreted onto dark matter (DM) halos 
where it gains gravitational energy. The collapsed gas is shock-heated to 
approximately the halo virial temperature, but radiative cooling 
in the inner (most dense) regions will lead to the formation cold
gas streams that sink into the center of the potential where the gas 
is transformed into stars (e.g., Maller \& Bullock 2004; Fukugita \& Peebles 2006).
Therefore, the CGM around present-day spiral galaxies is believed to represent a 
substantial gas reservoir from which galaxies acquire baryons to 
fuel star formation.

This simple concept of galaxy formation through gas accretion is altered by
the various types of feedback from active galactic nuclei (AGN), supernovae (SNe), 
or massive stellar winds, which deposit kinetic energy and chemically enriched material
into the CGM (e.g., Strickland et al. 2004; T\"ullmann et al. 2006). Also major and minor 
galaxy mergers can transport large amounts of cool and warm gas into 
the galaxies' circumgalactic environment (e.g., Yun et al.\,1994; Richter et al.\,2018).
As a result, the CGM is extremely multi-phase with cool and warm ($T=10^2-10^5$ K) 
gas streams being embedded in hot, virialized gas halos at $T=10^6-10^7$ K, typically.
Such hot gas halos often are referred to as ``galactic coronae'' (Spitzer 1956), in analogy 
to the Sun's hot coronal gas envelope. Hydrodynamcial simulations of cosmological galaxy 
formation, that include the necessary physics and that have the necessary spatial resolution 
(e.g., van\,de Voort et al.\,2018; Hani et al.\,2019), support the above outlined complex picture
of the CGM and its different phases. 

Observing the low-redshift CGM in all its phases requires the involvement of ground-based and
space-based telescopes in various wavelength ranges. Extended, cooler ($T<10^5$ K), 
predominatly neutral gas structures in the CGM, that originate in merger events
or in cooling accretion streams, may be observed using deep radio observations in the 
H\,{\sc i} 21cm line. Recent 21cm surveys indicate, however, that the cross section
of circumgalactic H\,{\sc i} emission features in external galaxies 
is very small (Pisano et al.\,2007) 
and most of the detected circumgalactic 21cm H\,{\sc i} streams appear to be related to 
galaxy mergers (e.g., Haynes et al.\,2011).

UV absorption spectroscopy of background AGN
is a powerful method to study H\,{\sc i} and metal-ion absorption of cold and
warm gas in the CGM of foreground galaxies, as the UV range
covers a large number of diagnostic transitions from low,
intermediate and high ions of heavy elements and the Lyman series
of neutral hydrogen. Over the last two decades, in particular, a large number
of absorption-line studies using UV spectral data from the 
Space Telescope Imaging Spectrograph (STIS) and the Cosmic
Origins Spectrograph, both intruments being installed 
on the Hubble Space Telescope (HST), have substantially improved our 
understanding of the nature of the diffuse circumgalactic gas component
of galaxies
(e.g., Borthakur et al.\,2016; Burchett et al.\,2019; Liang \& Chen 2014; 
Muzahid et al.\,2017; Prochaska et al. 2011, 2019; Richter et al.\,2017, 2018; 
Prochaska et al. 2011; Tumlinson et al. 2013; Wakker \& Savage 2009; Werk et al. 2013).
These (and other) studies have unveiled a large complexity in the circulation processes of 
metal-enriched gas around galaxies, governed by infall, gas accretion, major
and minor mergers, and outflows (see Tumlinson et al.\,2017 for a recent review).

From all CGM gas phases, the shock-heated, hot ($T>10^6$ K) phase is particularly 
difficult to be observed, owing to the very low density (log $n_{\rm H}=-2$ to $-5$, typically) 
of the gas and its high degree of ionization. X-ray continuum emission from the hot coronal
plasma of external galaxies has been studied using different instruments
(e.g., Bregman \& Houck 1997; O'Sullivan et al.\,2003; Strickland et al.\,2004; 
T\"ullmann et al.\,2006; Li et al.\,2008, 2016; Anderson \& Bregman 2010, 2011; Anderson et al.\,2016),
indicating that the coronae of Milky-Way type galaxies contain $\sim 10^{10}-10^{11} M_{\sun}$ of
gas, typically, exceeding the baryonic mass contribution of the cooler 
CGM phases by almost two orders of magnitude (see also Richter 2017).
For the Milky Way, also the X-ray lines of highly-ionized oxygen, 
O\,{\sc vii} and O\,{\sc viii}, represent important tracers of hot, circumgalactic
gas, as they can be observed either in absorption against X-ray bright AGN
or in emission (e.g., Paerels \& Kahn 2003; McCammon et al.\,2002; 
Nicastro et al.\,2002; Wang et al.\,2005; Williams et al.\,2005; 
Fang et al.\,2006; Miller \& Bregman 2013, 2015; Hodges-Kluck, Miller \& Bregman 2016;
Li \& Bregman 2017). These studies suggest that the hot CGM of the Milky Way has a 
total mass of $\sim 2-5\times 10^{10} M_{\sun}$ within $250$ kpc and that
the coronal gas co-rotates with the disk. 

Despite the overall importance and substantial baryon budget of million-degree 
coronal gas around galaxies, observational data in the X-ray band are still very limited 
(as it takes substantial effort to get them). Unfortunately, there are no strong resonance 
lines from high metal ions available in the UV/optical regime that would directly trace
million-degree gas in the CGM and IGM at $z=0$. York \& Cowie (1983) and Richter et al.\,(2014)
discussed the possibility to use the optical intersystem lines of 
[Fe\,{\sc x}] $\lambda 6374.5$  and [Fe\,{\sc xiv}] $\lambda 5302.9$ to
sample shock-heated, hot gas in the CGM and IGM. However, because
of the very small oscillator strengths of these 
forbidden transitions, an extremely high S/N of a few thousand would
be required to detect these lines in the spectra of background AGN,
which is currently not feasible.
As a consequence, our understanding of the physical nature
and spatial distribution of the hot CGM, in particular in the outer
halo near the virial radius, remains highly incomplete.
In this study, we explore the possibility to use thermally broadened 
H\,{\sc i} Ly\,$\alpha$ absorption lines (CBLAs) as tracers for the 
hot CGM around galaxies. 

The paper is organized as follows. In Sect.\,2,
we discuss the general motivation for using CBLAs as tracers of the hot gas 
distribution around low-redshift galaxies. In Sect.\,3, we present
in detail the setup of our semi-analytical model. The expected properties 
of CBLAs, as derived from our model, are discussed in Sect.\,4. 
In Sect.\,5, we provide four examples of CBLA candidates in archival 
UV data from the {\it Cosmic Origins Spectrograph} (COS) and 
the {\it Space Telescope Imaging Spectrograph} (STIS; both installed
on the Hubble Space Telescope, HST) and compare their properties 
to the model predictions. We discuss and summarize our results
in Sect.\,6. Supplementary equations, figures, and tables are provided in the 
Appendix.


\begin{figure}[t!]
\epsscale{0.95}
\plotone{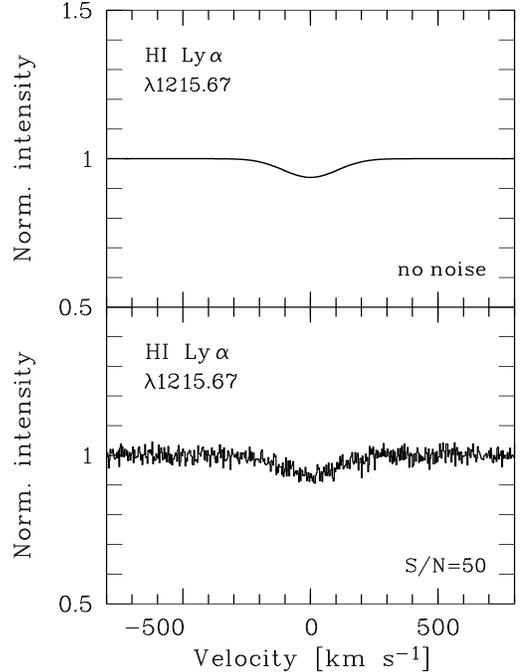}
\caption{
Example for a synthetic CBLA line that traces hot (million-degree) coronal
gas in the halo of an $L^{\star}$ galaxy along a sightline with
an impact parameter of $D=100$ kpc. The H\,{\sc i} column density 
is log $N$(H\,{\sc i}$)=12.9$ and the Doppler parameter is $b=105$ km\,s$^{-1}$.
The upper panel shows the synthetic Ly\,$\alpha$ line without noise, the lower panel shows
the same line at a S/N per pixel of 50 (corresponding to a S/N of $\sim 123$ per
$19$ km\,s$^{-1}$ wide resolution element, similar as for HST/COS).
}
\end{figure}

%

\begin{figure*}[t!]
\epsscale{1.00}
\plotone{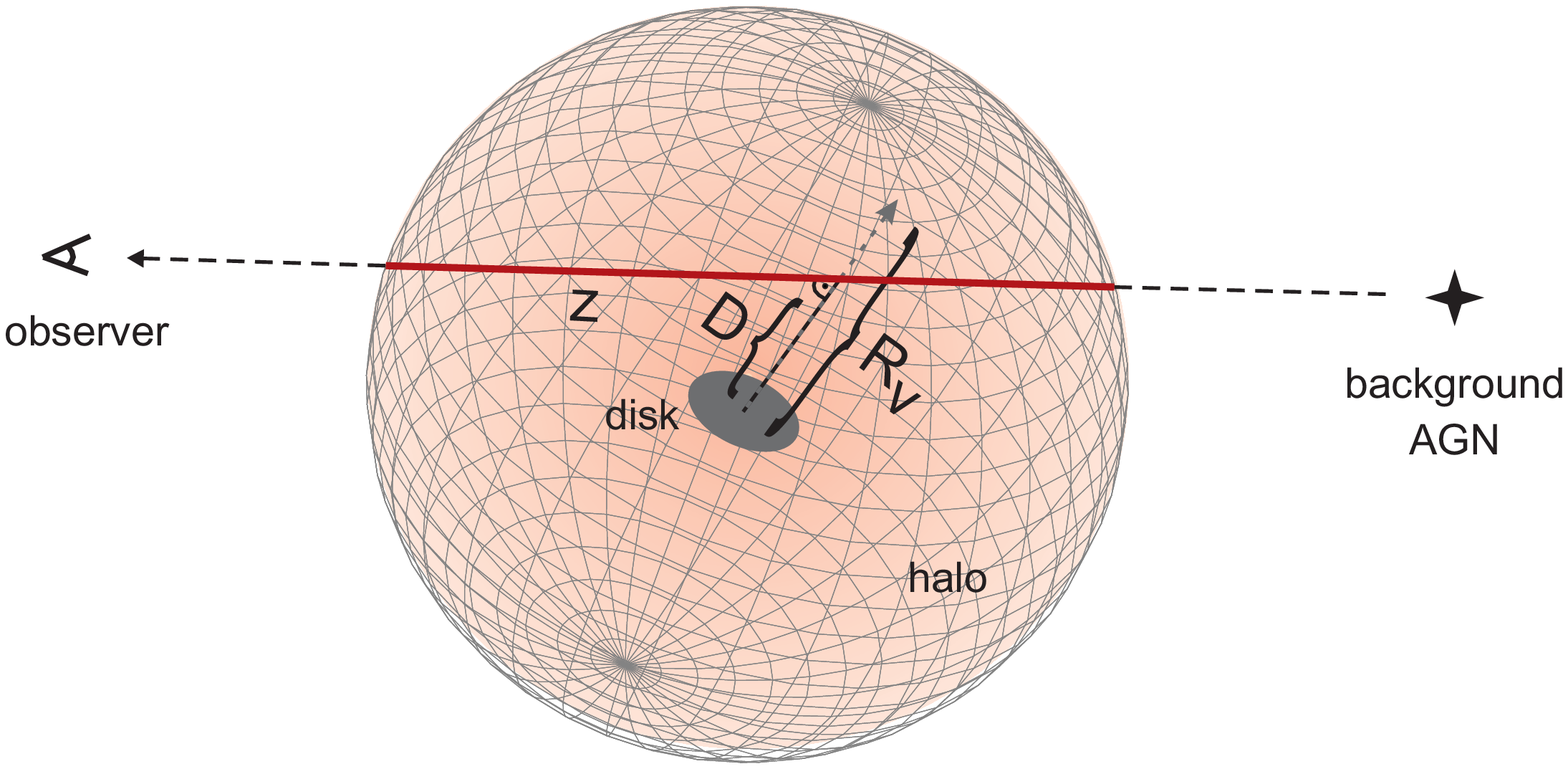}
\caption{Illustration of our modeling approach of CBLAs using the
{\tt halopath} code (Richter 2012). For a given galaxy
halo with virial halo mass, $M_V$, the radial density profile of the coronal gas (light-red
area) is calculated out to the virial radius, $R_V$. For a LOS
that passes the galaxy at impact parameter $D\leq R_V$, the H\,{\sc i} column
density is calculated from integrating the neutral gas density along the
LOS, while the thermal line broadening ($b_{\rm th}$) is determined from
temperature distribution along the LOS.
}
\end{figure*}


\section{Coronal Broad Ly\,$\alpha$ Absorbers}

Following galaxy-formation theories (e.g., White \& Frenk 1991), hot coronal 
gas around galaxies is expected to have temperatures close to the virial
temperatures of the galaxy halos, typically a few $10^6$ K.
At such high temperatures, the gas is almost fully ionized by collisions, 
with neutral hydrogen fractions $<10^{-5}$.
Photoionization from radiation originating in the UV background and
in the host galaxies themselves is mostly relevant for cooler gas with 
$T<10^6$ K (see Richter et al.\,2008).

Although the neutral gas fraction in such gas is tiny, there exists a sufficient 
number of neutral hydrogen atoms along a sightline that 
passes through the hot halo of a Milky-Way type galaxy to create a 
detectable Ly\,$\alpha$ absorption signal. The resulting so-called 
Broad Ly\,$\alpha$ Absorber (BLA) is shallow and broad, owing to the
substantial thermal line broadening caused by the high gas temperature.
In the following, we denote BLAs that trace the hot coronal gas of
galaxies as {\it Coronal Broad Ly\,$\alpha$ Absorbers} (CBLAs).
In anticipation of our modeling results, we show in Fig.\,1, as an example, 
the spectral appearance
of a CBLA that passes the hot halo of an $L^{\star}$ galaxy at an
impact parameter of $D=100$ kpc.

BLAs have been previously studied by us and other research groups to trace
the missing baryons in the Warm-Hot Intergalactic Medium (WHIM; Richter et al.\,2004, 
2006; Prause et al.\,2007; Danforth et al.\,2016; Savage et al.\,2011;
Narayanan, Savage \& Wakker 2012). Several dozen high S/N BLA candidate systems have been
detected so far, implying that they arise in shock-heated gas in the most
massive collapsing cosmological filaments.
Hydrodynamic cosmological simulations indicate, however, that the interpretation
of BLAs as tracers of the WHIM is afflicted with large systematic uncertainties, 
owing to the fact that large-scale gas flows and other non-thermal broadening
mechanisms contribute to the observed large BLA line widths (Richter, Fang \& Bryan 2006; 
Tepper-Garc\'ia et al.\,2012). Also instrumental effects, such as unresolved
multi-component structures, noise features, fixed-pattern artifacts as well
as continuum undulations intrinsic to the AGN's spectral energy distribution
limit the diagnostic power of broad Ly\,$\alpha$ features for the analysis of the WHIM.

BLAs that possibly are associated with warm/hot gas in the halos of individual galaxies
(i.e., CBLA candidates) have been reported regularly in previous studies
(e.g., Savage et al.\,2014; Stocke et al.\,2014; Johnson et al.\,2017), but a 
{\it systematic} investigation on how such broad absorbers might be related to the
hot coronal gas around their host galaxies has not been published so far. 
With this study, we are aiming at filling this gap.

The major advantages of analyzing circumgalactic CBLAs compared to intergalactic BLAs are
that i) we know exactly where we should look for them, namely along sightlines that pass 
galaxies within their virial radii at radial velocities defined by these galaxies,
ii) the hot gas is confined in a much smaller volume (i.e., within the virial 
radius of the galaxies), eliminating large-scale gas flows as line-broadening
mechanism, and iii) the temperature (and thus the ionization fraction) of the collisionally
ionized gas is expected to scale with galaxies' virial mass, allowing us to {\it predict}
the CBLA absorption properties for each individual galaxy/sightline pair.
However, the CGM is multi-phase, and therefore the cooler (less ionized) gas phases will
dominate the H\,{\sc i} optical depth in most CGM absorbers.
As a result, most CBLAs are expected to be embedded in (or hidden by) complex, 
multi-component H\,{\sc i} Ly\,$\alpha$ absorption systems.
This aspect will be further discussed in Sect.\,5, where we compare the model predictions
with UV spectral data from HST/COS and HST/STIS.

%

\section{Semi-analytic modeling of CBLAs}

\subsection{Model setup}

In the following, we outline our strategy for modeling the expected spectral
shape of CBLAs as a function of halo mass and sightline impact parameter.
Throughout the paper we adopt a standard $\Lambda$CDM cosmology with parameters
$\Omega_{\Lambda}=0.7$, $\Omega_{\rm m}=0.3$ and $H_0=70$ km\,s$^{-1}$\,Mpc$^{-1}$.

We assume that the hot halo gas is confined in a DM halo that is characterized by 
a Navarro-Frenk-White (NFW) density profile (Navarro, Frenk \& White 1995;
Klypin et al.\,2001). After the initial collapse, the gas is shock-heated to 
the temperature of the virialized halo, but will cool in the inner regions 
(within a characteristic cooling radius, $R_{\rm c}$) to become multi-phase.

We use the formalism developed by Maller \& Bullock (2004; hereafter MB04), which
provides analytic equations for the radial density and temperature profiles of the
residual hydrostatic hot gas halo in a NFW potential assuming gas cooling and
fragmentation under realistic conditions. With $r$ as radial variable, 
$M_V$ as virial halo mass, and $R_V$ as virial radius, the MB04 formalism
therefore provides 

\begin{equation}
n_{\rm H}(M_V,r)
\hspace{0.5cm}{\rm and}\hspace{0.5cm}
T(M_V,r)
\end{equation}

for $r\leq R_V$. The detailed equations for $n_{\rm H}(M_V,r)$ and $T(M_V,r)$ and their 
derivations (from the MB04 paper) are summarized in the Appendix (equations (A1)-(A12)).

We have implemented the equations for $n_{\rm H}(M_V,r)$ and $T(M_V,r)$ (equations
(A11) and (A12)) in our numerial {\tt halopath} code, a code developed to model
the absorption properties of multi-phase halos of galaxies in different mass ranges 
(Richter 2012). 

Theoretical studies and simulations imply that only massive halos
(log $M/M_{\sun}\geq 11.3$) are expected to develop collisionally ionized,
coronal gas halos from gravitational collapse (e.g., Gutcke et al.\,2017).
However, observations suggest that also lower-mass galaxies are surrounded
by warm/hot gas (e.g., Johnson et al.\,2017), possibly generated and maintained by
winds and outflows. Such dwarf galaxies also might give rise to CBLAs
and could substantially contribute to the cosmological CBLA cross section at $z=0$.

On the high-mass end, galaxies with masses log $M/M_{\sun}\geq 12.5$ are rare
and thus the cosmological cross section of their halos is small. In addition,
the neutral gas fraction in the coronae of such massive galaxies 
are expected to be very small with extremely large thermal line widths for
the resulting Ly\,$\alpha$ absorption, so that no detectable CBLA signal is expected
to emerge from such halos at realistic S/N ratios ($\leq 100$ per resolution element).

Based on these considerations, we have created a set of model 
halos with virial halo masses in the 
for us relevant range log $(M_V/M_{\sun})=10.6-12.6$ and in steps of $0.2$ dex.
Each model halo is characterized by a radial grid of data points at 1 kpc 
resolution that reaches up to the virial radius.

The {\it neutral} hydrogen volume density, $n_{\rm HI}$, in the coronal gas
at radius $r$ is given by the relation

\begin{equation}
n_{\rm HI}(M_V,r)=f_{\rm HI}(T)n_{\rm H}(M_V,r),
\end{equation}

where $f_{\rm HI}(T)$ denotes the neutral gas fraction.
In a collisional ionization equilibrium (CIE), the neutral gas fraction
in a plasma depends only on the gas temperature (i.e., it is density-independent).
Following our initial work on BLAs (Richter et al.\,2004), $f_{\rm HI}(T)$ can 
be expressed with a polynomial in the form

\begin{equation}
{\rm log}\,f_{\rm HI}(T)=13.9-5.4\,{\rm log}\,T+0.33\,{\rm log}\,T^2.
\end{equation}

Since $n_{\rm H}$ and $T$ is pre-defined in each grid point from the MB04 coronal gas model,
$n_{\rm HI}$ can be calculated in each point using Eq. (2) and (3).

To calculate the neutral hydrogen column density of a CBLA, $N$(H\,{\sc i}), along a halo 
sightline at impact parameter $D$, we need to integrate
$n_{\rm HI}(r)$ along the path through the coronal gas distribution.
With $z$ being the spatial coordinate along the line of sight (LOS), the 
integral formally can be written as

\begin{equation}
N({\rm HI})\big\vert_D=\int\limits_{-\infty}^{+\infty}n_{\rm HI}
\left(\sqrt{D^2+z^2}\right){\rm d}z =
2\int\limits_D^{+\infty}\frac{r\,n_{\rm HI}(r)\,{\rm d}r}{\sqrt{r^2-D^2}},
\end{equation}

where we use a transformation of the integration variable in the form
$r=\sqrt{D^2+z^2}$. In reality, we obtain $N$(H\,{\sc i}) in our 
{\tt halopath} model halos by numerically integrating $n_{\rm HI}$ 
over all grid cells along the halo sightline. In Fig.\,2 we sketch
the geometric setup of our modeling approach and indicate the 
parameters involved. 

The intrinsic width of an the resulting $N$(H\,{\sc i}) Ly\,$\alpha$
absorption line (i.e., the CBLA) is characterized by its 
Doppler parameter/$b$-value, which is composed of 
a thermal ($b_{\rm th}$) and a non-thermal ($b_{\rm non-th}$) component:

\begin{equation}
b=\sqrt{b_{\rm th}^{\,2}+b_{\rm non-th}^{\,2}}.
\end{equation}
 
In our idealized model, we assume that in the case of CBLAs, thermal motions of the 
coronal gas particles dominate over other broadening mechanisms.
Because $b_{\rm th}$ and $b_{\rm non-th}$ are added quadratically, we 
therefore ignore any contributions from non-thermal motions and
assume

\begin{equation}
b=b_{\rm th}=\sqrt{\frac{2k\langle T \rangle}{m_{\rm H}}},
\end{equation}

where $m_{\rm H}$ is the mass of a hydrogen atom and 
$\langle T \rangle$ is the mean coronal gas temperature 
along the sightline. 
Non-thermal motions of hot gas could be relevant for galactic
outflows and merger events, which should be kept in mind
when it comes to the interpretation of observed line-widths
in CBLA candidate systems.

Note that in the MB04 model the coronal gas is {\it not} isothermal
(because of cooling; see Appendix, Eq. A10). 
Each halo sightline passes the gas as at different radii, thus at different temperatures. 
We calculate $\langle T \rangle$ by taking the density-weighted
mean of the gas temperature in all grid points along the LOS. 

In summary, the model provides $N$(H\,{\sc i}) and $b$ for a CBLA
as a function of halo mass and impact parameter. It 
thus allows us to generate synthetic spectra of CBLAs for any 
given galaxy-halo sightline parametrized by $(M_V, D)$.
An example of such a synthetic spectrum is shown in Fig.\,1,
generated by the {\tt fitlyman} software implemented in
ESO/MIDAS (Fontana \& Ballester 1996).

The model also allows us to systematically investigate the distribution
of column densities and $b$ values for our set of MB04 model halos,
as will be discussed in the following section.
For comparison, we discuss in the Appendix the CBLA
properties for an alternative, isothermal halo model.

It is important to keep in mind that this semi-analytic approach describes, 
by construction, an (over)idealized circumgalactic gas environment. Our model cannot 
take into account other important aspects of galaxy formation and evolution
(e.g., intrinsic gas-density and -temperature variations, non-spherical
halo geometries, minor and major mergers, feedback processes, 
cosmological environment, etc.)
that potentially influence the strength and shape of broad Ly\,$\alpha$ 
absorption arising in hot coronal gas in realistic galaxy environments.
These aspects will be studied by us in a future paper, where we will use
high-resolution CGM simulations to explore the spectral signatures of
million-degree gas around galaxies.


\begin{figure}[t!]
\epsscale{0.90}
\plotone{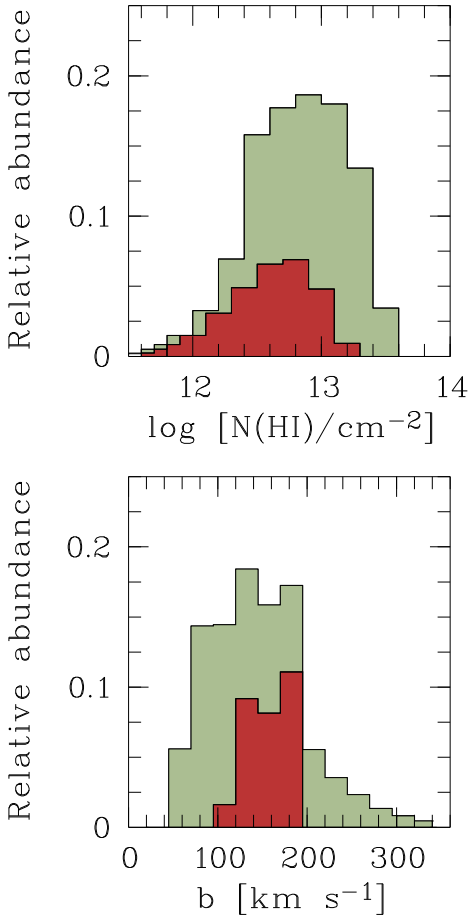}
\caption{Distribution of CBLA H\,{\sc i} column densities (upper panel)
and $b$ values (lower panel) for the local galaxy population, as predicted
from our model. The red-shaded insets show the distributions for the
coronal gas residing outside the cooling radius (Sect.\,4.1).
}
\end{figure}


\begin{figure}[t!]
\epsscale{1.00}
\plotone{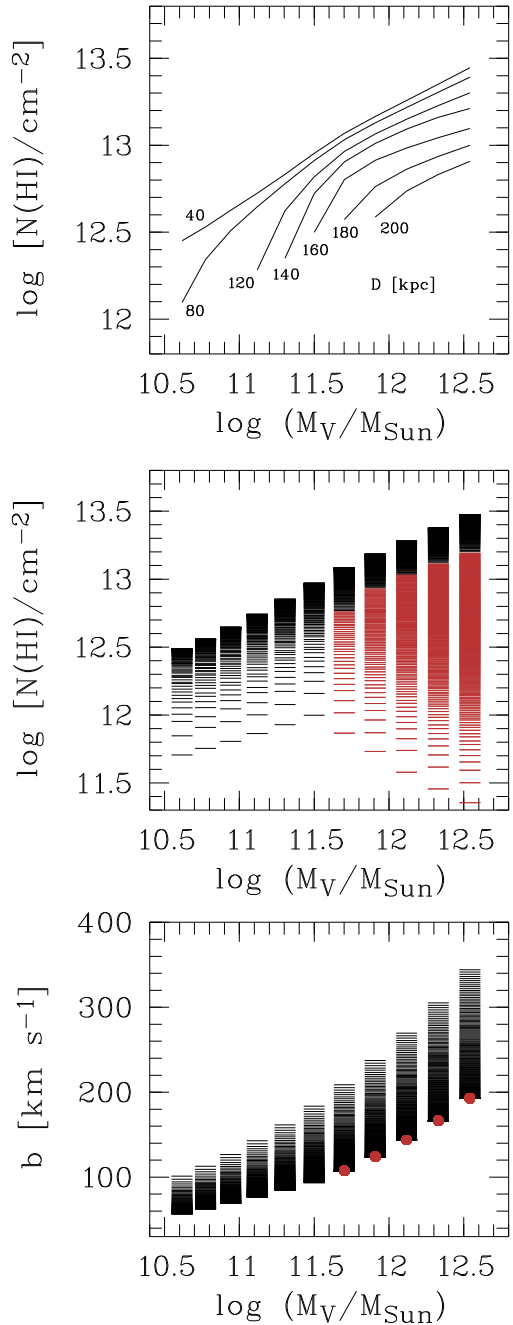}
\caption{Dependence of CBLA H\,{\sc i} column densities and $b$ values on
the virial halo mass and the LOS impact parameter. 
{\it Upper panel:} log $N$(H\,{\sc i}) vs. log $M_V$ for constant impact
parameters (black solid lines, lables for D in kpc).
{\it Middle panel:} log $N$(H\,{\sc i}) vs. log $M_V$ for the full range
of impact parameters, $D\leq R_V$, with the red bars indicating LOS beyond
the cooling radius.
{\it Lower panel:} $b$(H\,{\sc i}) vs. log $M_V$ for the full range
of impact parameters.
}
\end{figure}


\section{Properties of CBLAs}

\subsection{Distribution of H\,{\sc i} column densities and $b$ values}

To investigate the statistical properties of the CBLAs, we have
generated 2134 lines of sight passing through eleven MB04 model halos at $z=0$ in the 
mass range log $(M_V/M_{\sun})=10.6-12.6$ (corresponding to galaxy luminosities
in the range $0.1-10\,L^{\star}$) at impact 
parameters $0\leq D\leq R_V$ (in steps of 1 kpc). 
Note that the virial radius $R_V$ scales with $M_V$ as given 
in Eq.\,(A1).

In the upper panel of Fig.\,3, we show the distribution of logarithmic H\,{\sc i} column 
densities for all these 2134 sightlines (green-shaded area) in bins of 0.2 dex.
All H\,{\sc i} column densities lie in the range log $N$(H\,{\sc i}$)=11.3-13.5$, 
with 84 percent of the absorbers having log $N$(H\,{\sc i}$)=12.4-13.4$.
The distribution peaks at log $N$(H\,{\sc i}$)=12.9$, a value that can be regarded
as ``characteristic'' for CBLAs.
The red-shaded area displays the H\,{\sc i} distribution of a sub-sample of 645 CBLAs
that trace the outer hot halos beyond the cooling radius (i.e., $r>R_{\rm c}$, Eq.\,A10).
Here, the column densities are generally smaller compared to central sightlines due
to the lower gas densities in the outer halos and the 
shorter absorption pathlengths at larger $D$. This effect also explains the wing in the 
distribution at log $N$(H\,{\sc i}$)\leq 12.4$.

In the lower panel of Fig.\,3, we show the distribution of H\,{\sc i} $b$ values  
for the total sample of 2134 absorbers (green-shaded area) and the CBLA subsample
with $r>R_{\rm c}$ (red-shaded area). For the total sample, all $b$ values are
between 50 and 350 km\,s$^{-1}$, with 82 percent of the absorbers having 
$b=70-200$ km\,s$^{-1}$. Here, the distribution peaks at a characteristic 
value of $b\approx 140$ km\,s$^{-1}$. The $b$-value distribution for the 
sub-sample does not extend beyond 
$200$ km\,s$^{-1}$. This is because the $b$ values $>200$ km\,s$^{-1}$ 
originate in absorbers in the {\it inner} halos at $r\leq R_{\rm c}$,
where the CGM is assumed to be multiphase due to enhanced cooling. Following MB04, 
the residual hot gas in the inner halo will change its pressure adiabatically to 
adjust to a hydrostatic eqilibrium. This will lead to a temperature 
{\it increase} for the inner hot halo at $r\leq R_{\rm c}$, with $T(r)$ 
increasing for decreasing $r$ with by a factor of 
a few compared to the initial hot halo temperature at $r> R_{\rm c}$,
where the gas is assumed to be isothermal.
As a consequence, CBLAs arising at small impact parameters trace hotter
gas than those at larger impact parameters, leading to particularly broad
lines with large $b$ values. 


\begin{figure}[t!]
\epsscale{1.00}
\plotone{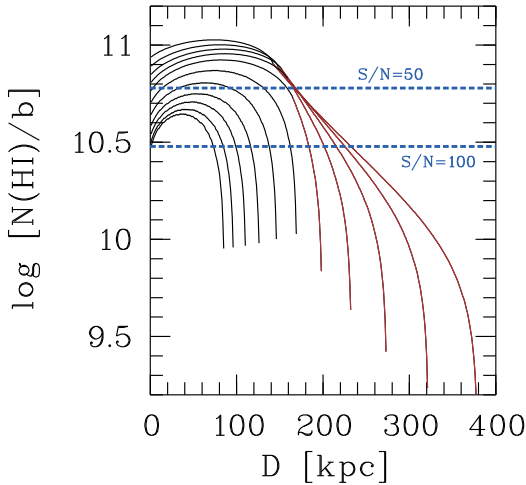}
\caption{Distribution of log $(N/b)$ (CBLA detectability criterion; see
Sect.\,4.2) as a function of impact parameter for the different viral halo 
masses in our model (log $M_V=10.6-12.6$ from left to right, in steps of 0.2 dex).
The observational $(N/b)$ detection limits for a S/N (per resolution element) 
of $100$ and $50$ are overlaid in blue. 
}
\end{figure}


In Fig.\,4, we show the distributions of H\,{\sc i} column densities and
$b$ values as function of halo mass, $M_V$. In the upper panel, we display 
lines of constant impact paramter in the $M_V$/$N$(H\,{\sc i}) parameter
space. This figure shows that only the central ($D\leq 100$ kpc) sightlines 
of massive (log $M_V\geq 11.6$) galaxies produce CBLAs with substantial 
H\,{\sc i} column densites above log $N$(H\,{\sc i}$)=12.8$. The middle and 
lower panel in Fig.\,4 show for each model halo what range in H\,{\sc i} 
column density and $b$ value is covered by CBLAs, when the impact
parameter is varied from $0<D\leq R_V$ (in steps of 1 kpc, horizontal
bars). The red bars indicate sightlines passing the outer halo beyond
the cooling radius (i.e., $D>R_{\rm c}$). Fig.\,4 may be used to 
predict the strength of a CBLA for a given halo sightline with known
impact parameter and galaxy luminosity (from which $M_V$ can be estimated).
A table, that lists the expected values for log $N$(H\,{\sc i}) and $b$(H\,{\sc i}) 
for all values of $M_V$ and $D$, can be made available on request.


\begin{deluxetable*}{rrrrrrrr}
\tabletypesize{\scriptsize}
\tablewidth{0pt}
\tablecaption{Coronal gas masses}
\tablehead{
\colhead{No.} &
\colhead{log $M_V\,^{\rm a}$} &
\colhead{$L/L^{\star}\,^{\rm b}$} &
\colhead{$R_V\,^{\rm c}$} &
\colhead{$R_{\rm c}\,^{\rm d}$} &
\colhead{log $M_{\rm b}\,^{\rm e}$} &
\colhead{log $M_{\rm cor}\,^{\rm f}$} &
\colhead{$f_{\rm cor}\,^{\rm g}$} \\
\colhead{} & \colhead{} & \colhead{} & \colhead{[kpc]} & \colhead{[kpc]} &
\colhead{} & \colhead{} & \colhead{}
}
\startdata
 1 & 10.62 &  0.10 &   86   &  86  &  9.85  &   8.27 & 0.03 \\
 2 & 10.77 &  0.16 &   97   &  97  & 10.00  &   8.61 & 0.04 \\
 3 & 10.94 &  0.25 &  111   & 111  & 10.17  &   8.97 & 0.06 \\
 4 & 11.12 &  0.40 &  127   & 127  & 10.35  &   9.33 & 0.09 \\
 5 & 11.31 &  0.63 &  147   & 147  & 10.54  &   9.72 & 0.15 \\
 6 & 11.50 &  1.00 &  170   & 170  & 10.73  &  10.11 & 0.24 \\
 7 & 11.70 &  1.58 &  199   & 164  & 10.93  &  10.48 & 0.35 \\
 8 & 11.91 &  2.51 &  233   & 159  & 11.14  &  10.81 & 0.46 \\
 9 & 12.11 &  3.98 &  274   & 153  & 11.35  &  11.11 & 0.58 \\
10 & 12.33 &  6.31 &  322   & 147  & 11.56  &  11.40 & 0.69 \\
11 & 12.54 & 10.00 &  379   & 142  & 11.77  &  11.67 & 0.79 \\
\enddata
\tablecomments{
$^{\rm a}$\,Virial halo mass, in solar units;
$^{\rm b}$\,galaxy luminosity;
$^{\rm c}$\,virial radius;
$^{\rm d}$\,cooling radius (see Appendix, Eq.\,A10);
$^{\rm e}$\,total baryon mass, in solar units;
$^{\rm f}$\,coronal baryon mass, in solar units;
$^{\rm g}$\,coronal baryonic mass fraction.
}
\end{deluxetable*}


\subsection{On the detectability of CBLAs}

Figs.\,3 and 4 indicate that CBLAs span a broad range in 
$N$(H\,{\sc i}) and $b$. Detecting broad, shallow absorption features 
in UV data with limited S/N is challenging, however. The detection 
significance depends on both the depth and the width of the absorption as well
as on the local S/N. In our previous study (Richter et al.\,2006), we have defined an
empirical criterion for the detectability of a BLA in the form:

\begin{equation}
\left[\frac{N({\rm HI})}{{\rm cm}^{-2}}\right]\left[\frac{b({\rm HI})}{{\rm km\,s}^{-1}}\right]^{-1}
\geq \frac{3\times 10^{12}}{\rm (S/N)_{\rm res}}.
\end{equation}

Here, (S/N)$_{\rm res}$ is the local S/N per resolution element.
For a ``typical'' CBLA having log $N$(H\,{\sc i}$)\approx 13$ and
$b$(H\,{\sc i}$)\approx 130$ km\,s$^{-1}$ (log $(N/b)=10.9$), this implies that a 
S/N of $\approx 40$ per resolution element is required to securely detect
such an absorber. This value corresponds to a central absorption depth 
of ${\cal D}=0.06$ in the CBLA line (see Appendix, Eq.\,A16). 
Combining the statistics for $N$ and $b$, we find that 
80 percent of the CBLAs have log $(N/b)\geq 10.5$. Since for
each halo, $N$(H\,{\sc i}) (and thus $(N/b)$) decreases with increasing 
impact parameter due to the decreasing path length through the halo, 
CBLAs that trace the isothermal, hot halo component at $r\geq R_{\rm c}$
are particularly difficult to detect. In Fig.\,5 we show, how log $(N/b)$
varies with $D$ for the eleven model halos and out to which impact parameter CBLAs 
can be detected at (S/N)$_{\rm res}=50$ and $100$ (blue horizontal lines).

Because the strongest CBLAs sample the inner regions of galaxy halos, 
they are expected to blend with narrow (and predominantly
stronger) H\,{\sc i} features stemming from the warm/cool ($T<10^5$ K)
CGM that traces infalling and outflowing gaseous material.
Therefore, many CBLAs may be hidden in multi-component H\,{\sc i} Ly\,$\alpha$ 
profiles and may not be readily visibile (and $(N/b)$ may not be a 
meaningful criterion for their detection).
In such cases, careful profile-fitting of H\,{\sc i} and associated
metal-ions that trace the $T<10^5$ K gas phases is required to search
for evidence of broad, shallow H\,{\sc i} components that 
might be related to hot, coronal gas components in multi-phase CGM absorbers.
Examples for such multi-phase CGM absorbers possibly containing a CBLA
will be presented in Sect.\,5.

\subsection{Hot gas mass traced by CBLAs}

The total hot gas mass traced by CBLAs can be determined by integrating 
for each model halo the individual coronal mass shells in our {\tt halopath}
code from inside out (from $r=0$ to $r=R_V$), where we assume an average  mass
per particle of $m_{\rm p}=1.4\,m_{\rm H}$, accounting for the presence of helium
and heavy elements in the gas.

In Table 1, we summarize for the different galaxy mass bins the resulting 
logarithmic coronal gas masses (7th column), the baryon fraction 
in the corona (8th column), as well as other model parameters 
(such as virial radius, expected galaxy luminosity, and cooling radius). 

The baryon fraction in the coronal gas increases with increasing mass up to
a value of $\sim 80$ percent. Therefore, these models underline that 
the hot CGM represents a major (eventually dominant) baryon reservoir in massive
galaxies that should be constrained by observations to test galaxy-formation
models.


\begin{figure}[t!]
\epsscale{0.92}
\plotone{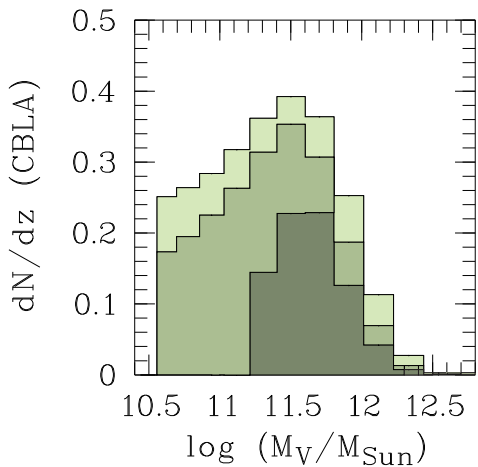}
\caption{Expected number density of CBLAs per unit redshift, $d{\cal N}/dz$,
for different virial halo-mass bins (Table 1, second  column) for different
sensitivity limits (light green=total sample; middle-dark green=S/N of 100;
dark green=S/N of 50).
}
\end{figure}


\subsection{Cosmological cross section of CBLAs}

If we consider all galaxies in the mass range log $(M/M_V)=10.6-12.6$ 
($(L/L^{\star}=0.1-10$) at $z=0$,
what would be the absorption cross section of CBLAs produced by their hot halos?

The absorption cross section of intervening absorbers usually is characterized
by the number density of absorbers per unit redshift, $d{\cal N}/dz$. The
expected number density of CGM absorbers depends on both the space density of 
galaxies, $\phi$, and the projected geometrical cross section, $A$, of the CGM 
phase traced (see Richter et al.\,2016, hereafter R16).
In case of CBLAs, which trace the hot halo gas out to the virial radius at
hundred percent covering fraction, the geometrical cross section is 
simply $A=\pi R_V^2$, so that the number density of absorbers per unit redshift
can be expressed as

\begin{equation}
\frac{d{\cal N}}{dz}({\rm CBLAs})=
\phi\,\pi R_V^2\,\frac{c\,(1+z)^2}{H(z)}.
\end{equation}

Here, $H(z)$ is the Hubble parameter, defined as
$H(z)=H_0\,(\Omega_{\rm m}\,(1+z)^3+\Omega_{\Lambda})^{1/2}$
(assuming a matter-dominated flat Universe with a cosmological constant).

Eq.\,(8) has to be considered for each galaxy mass bin, $\Delta M_V$, separately,
because the galaxy space density, $\phi$, and the virial radius, $R_V$, both are 
functions of $M_V$. In the Appendix, we show how $\phi (M_V)$ can be obtained
from the local galaxy luminosity function (B2) and how $L$ and $M_V$ are related (B3).
In Fig.\,6, the expected $d{\cal N}/dz$(CBLA) is show as a function of $M_V$ for the 
eleven mass bins in our galaxy model sample. The light green area indicates 
the distribution of $d{\cal N}/dz$(CBLA) for an infinte S/N, the intermediate
green area shows the distribution for a S/N per resolution element of 100, 
the dark green area assumes a S/N of 50. If we integrate over these distributions, we obtain 
total number densities per units redshift of
$d{\cal N}/dz$(CBLA)$_{\infty}=2.6$,
$d{\cal N}/dz$(CBLA)$_{100}=2.1$, and
$d{\cal N}/dz$(CBLA)$_{50}=0.8$
for the halos in the adopted mass range log $(M/M_V)=10.6-12.6$.


\begin{figure*}[t!]
\epsscale{1.00}
\plotone{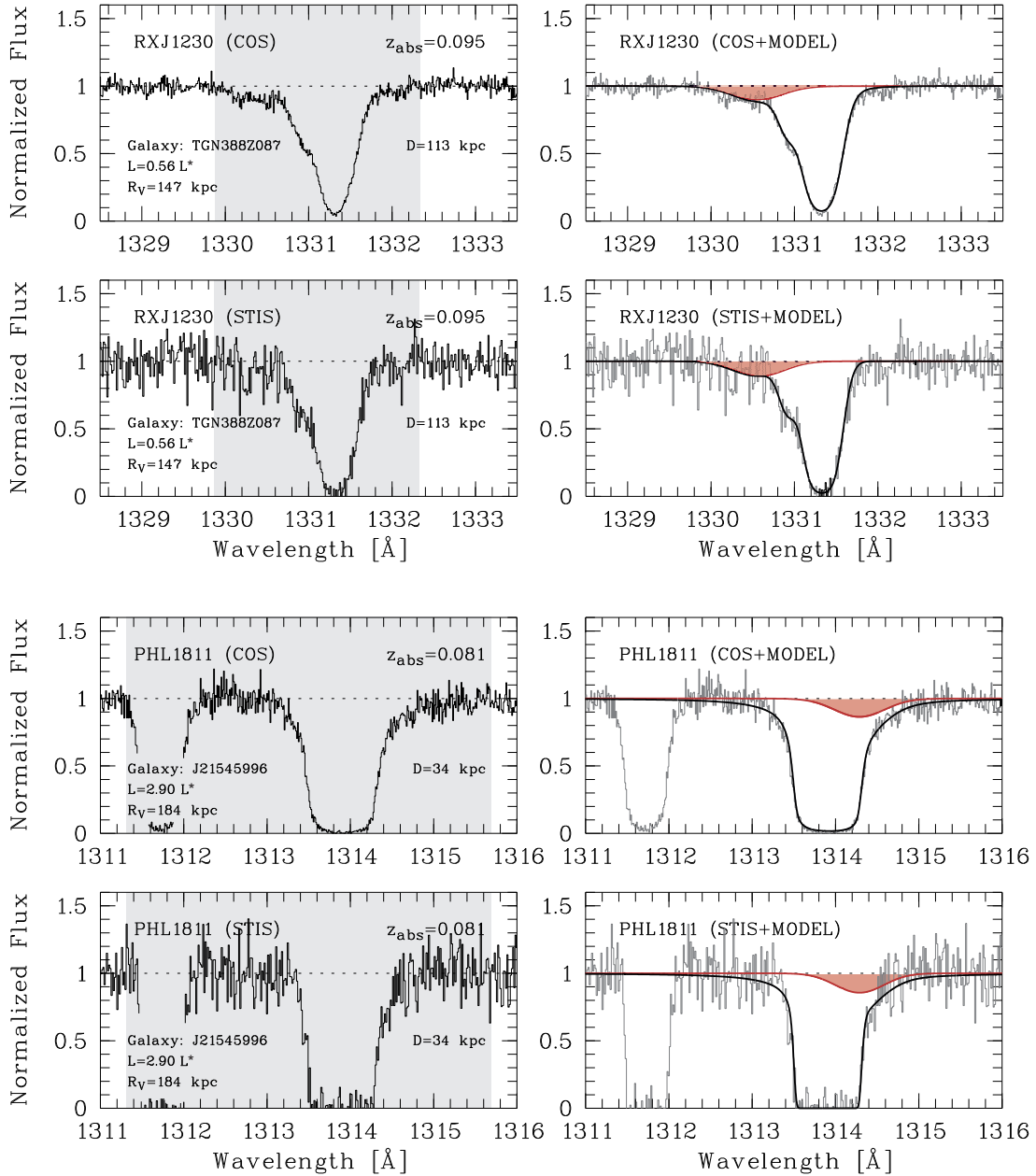}
\caption{CBLA candidate systems in the COS and STIS spectra of the AGN
RX\,J\,1230.8+0115 and PHL\,1811. The left panels show the raw spectral
data in the overall wavelength ranges where CBLA absorption in the halos of
intervening galaxies is expected. Galaxy data and impact parameters are
listed in the panels. The gray shaded areas indicate the expected
range for CBLA absorption based on the accuracy of the galaxy redshifts and
allowing for co-rotation of the coronal gas with the disk. The right panels
show the data together with the best-fitting multi-component models of the
Ly\,$\alpha$ absorbers (black solid line) and the modeled CBLA absorption
(red-shaded area).
}
\end{figure*}


\begin{figure*}[t!]
\epsscale{1.00}
\plotone{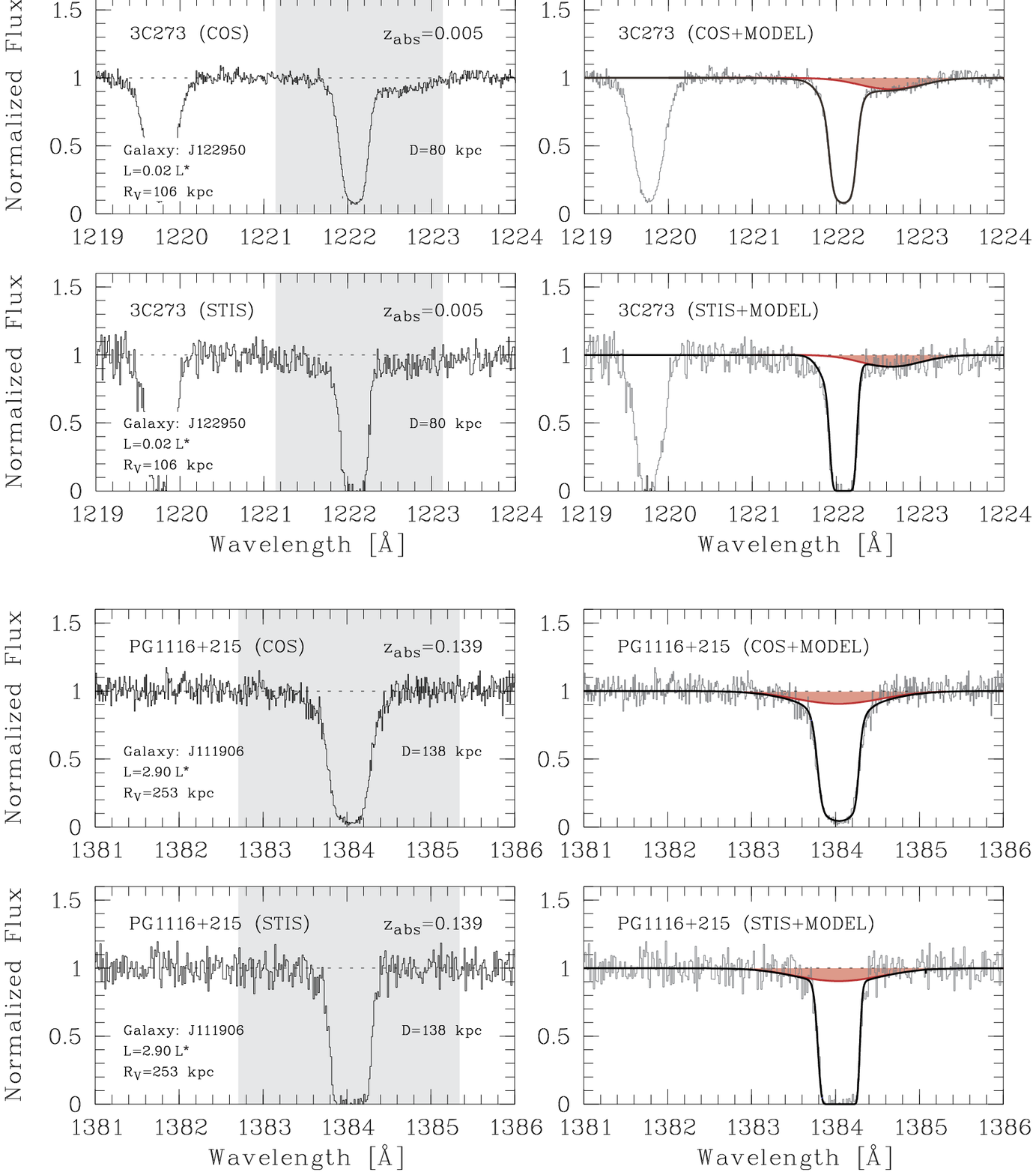}
\caption{Same as Fig.\,7, but for the LOS towards 3C\,273 and
PG\,1116+215.
}
\end{figure*}


\section{Confronting theory with observations}

\subsection{CBLA search strategy}

Our predictions about the strength, spectral shape, and frequency of CBLAs
at $z=0$ as potential tracers of shock-heated, hot halo gas can be
tested with existing UV absorption-line data. For this, we are particularly interested 
in QSO sightlines that are known to pass the halos of nearby massive galaxies
within their virial radii and for which good S/N UV spectral data 
are available.

We have searched for CBLA candidate systems in HST archival 
QSO data that we have used in previous studies (Richter et al.\,2009, 2016, 2017; Herenz et al.\,2013)
along galaxy halo sightlines that have been observed with {\it both} COS and STIS.
This allows us to assess the significance of CBLA candidates in two independent 
observational data sets from different HST instruments.
We used the galaxy data collected for our 2016 CGM
survey (R16) together with recent literature data 
(Stocke et al.\,2013; Keeney et al.\,2018)
and have compiled a list of galaxy/sightline pairs with impact 
parameters $D \leq R_V$ for galaxies with
known luminosities and with existing COS (STIS) spectra that have 
a S/N$>30$ ($>10$) per resolution element.

For the identification of a CBLA candidate system, we require simultaneous (but 
independent) evidence for a broad, shallow absorption feature in both the COS and STIS data
within $|\Delta v|=500$ km\,s$^{-1}$ of the redshift of the intervening galaxy. 
The COS data used here (G130M grating) have a spectral resolution of $\sim 19$ km\,s$^{-1}$ FWHM,
while the resolution of the STIS data (E230M grating) is $\sim 7$ km\,s$^{-1}$ FWHM.

As we will demonstrate below, the presence of broad, shallow Ly\,$\alpha$ absorption embedded in 
a multi-component CGM absorber is not always immediately evident from a by-eye inspection.
For most cases, it requires a careful modeling of the cooler H\,{\sc i} absorption
components (tracing the $T<10^5$ K CGM) to identify broad absorption components in the model residuals.
Even then, however, the fit/model solutions for complex multi-component CGM absorbers are never 
unique, as the modeling bases on a variety of assumptions. Most critical for the identification 
of shallow, broad features is the choice of the local QSO continuum, which may have local
undulations that are difficult to be accounted for. 
In addition, an apparently broad absorption component may be composed of several,
narrow components that are unresolved in the COS/STIS spectral data, in particular,
if the S/N is only moderate or low (see also Richter et al.\,2006, their Fig.\,1).
A standard procedure in the modeling of multi-component absorption-line systems is
to assign high metal ions (e.g., O\,{\sc vi}, C\,{\sc iv}) to broad H\,{\sc i} 
components and thereby tie the velocity-component structure for these species,
assuming that they are co-spatial. However, given the inhomogeneous distribution of 
the multiple gas phases in the CGM and a potentially non-uniform metallicity distribution in the gas,
this assumption my not be justified in general.

Notwithstanding these restrictions, we have started to re-model the spectral shape of 
multi-component Ly\,$\alpha$ CGM absorbers detected with STIS and COS data, all of them
being well-known multi-phase CGM absorption systems analyzed in previous studies.
For the component modeling we used Voigt profiles convolved with the 
appropriate STIS/COS line-spread functions. The modeling provides radial 
velocities (or redshifts), H\,{\sc i} column densities, and $b$ values for each 
absorption component (see Richter et al.\,2013 for a detailed description of 
the modeling code). 

In this paper, we present four typical CBLA candidate systems from this
search and discuss them in detail. The modeling results for these
four systems are summarized in Table 2.
The full survey of CBLAs will be presented in a forthcoming paper.

\subsection{The CBLA candidate at $z=0.095$ towards RX\,J\,1230.8+0115}

The sightline towards the Seyfert 1 galaxy RX\,J\,1230.8+0115
($z_{\rm em}=0.117, V=14.42$) passes the $0.56 L^{\star}$ galaxy 
2dFGRS-TGN388Z087 ($z=0.095$) at an impact parameter of $D=113$ kpc (Keeney et al.\,2018).
The COS data (S/N of $\sim 73$ per resolution element near 1330 \AA) and the 
STIS data (S/N of $\sim 11$ per resolution element near 1330 \AA) show a 
multi-component H\,{\sc i} Ly\,$\alpha$ absorber centered at 
$z_{\rm abs}=0.0951$; Fig.\,7, upper four panels). The visual inspection
of the COS and the STIS data (upper two left panels in Fig.\,7)
indicates that the dominating strong Ly\,$\alpha$ absorption component 
at $1331.3$ \AA\, is accompanied by at least two weak satellite components,
from which the very broad component near $1330.4$ \AA\, (our 
CBLA candidate) is readily visible in the COS data (while it is
hidden in the noise in the STIS data).

From a simultaneous fit/model of the COS and STIS data (Fig.\,7, upper right panels, 
black solid line) for the entire multi-component absorber at $z_{\rm abs}=0.0951$ we 
obtain for the CBLA component log $N$(H\,{\sc i}$)=13.14\pm 0.16$ and $b=90\pm20$ 
km\,s$^{-1}$. The preferred model for the CBLA candidate is indicated with the 
red-shaded area. Alternatively, it is also possible to fit this apparently broad component
with a series of three narrow components. Such a fit would better reproduce the individual
$1-2\sigma$ spikes in the COS/STIS noise patterns, but is, in view of the S/N, 
statistically not justified and thus arbitrary.
We conclude that the observed broad feature represents a convincing CBLA candidate.
Also Danforth et al.\,(2016) have identified this feature as a BLA. From a
fit of the COS data they obtain log $N$(H\,{\sc i}$)=12.99\pm 0.06$ and $b=70\pm10$,
thus in agreement with our results from the combined COS and STIS data.

For comparison: our CBLA models predicts log $N$(H\,{\sc i}$)=12.67$ and $b=89$ 
km\,s$^{-1}$ for $L\approx 0.6 L^{\star}$ and $D=113$ kpc. While the $b$ value from the model
agrees well with our fit, the H\,{\sc i} column density measured in the spectral data 
is susbtantially higher ($\sim 0.5$ dex) compared to what is predicted by the model. This 
discrepancy could be, for instance, related to an excess of hot gas in the extended halo of 
2dFGRS-TGN388Z087 (compared to the model), a contribution of H\,{\sc i} absorption from cooler 
gas phases, and/or non-equilibrium ionization conditions.
A general comparison between the measurements and the model predictions will be provided in
Sect.\,5.6.

\subsection{The CBLA candidate at $z=0.081$ towards PHL\,1811}

Another CBLA candidate is identified at $z_{\rm abs}=0.081$ along the 
line of sight towards the Seyfert 1 galaxy PHL\,1811 ($z_{\rm em}=0.194, V=16.80$), 
which passes the $2.90 L^{\star}$ galaxy 2MASS\,J21545996-0922249 at an impact 
parameter of $D=34$ kpc (Jenkins et al.\,2003, 2005).
This is a prominent CGM absorption system containing a strong 
Ly\,$\alpha$ absorption component centered at $1313.9$ \AA\,
that traces cooler gas in the halo of 2MASS\,J21545996-0922249 (Fig.\,7, lower four panels). 
The main Ly\,$\alpha$ component is accompanied by metal absorption 
in various low and intermediate ions (e.g., C\,{\sc ii}, C\,{\sc iv},
Si\,{\sc iii}, Si\,{\sc iv}; see Jenkins et al.\,2005, their Fig.\,1, for 
velocity profiles of the STIS data and R16, their Fig.\,A.1, for velocity profiles 
of the COS data). A weaker satellite component is seen bluewards of 
the main absorption component, but only in the lines of C\,{\sc iv} and Si\,{\sc iv} (R16).

It is the {\it red} wing of the Ly\,$\alpha$ profile, however, that exhibits
extended, shallow absorption that cannot be readily modelled with the 
component structure defined by the metal lines. Fitting the STIS data
(S/N is $\sim 9$ per resolution element near 1315 \AA)
alone, Jenkins et al.\,(2005) attribute the wing to a continuum undulation
(their Fig.\,2) on top the Ly\,$\alpha$ absorption. The much better COS data
(S/N of $\sim 41$ per resolution element near 1315 \AA; see R16, their Fig.\,1), 
indicates, however, that the shallow red wing more likely is attributed to 
an additional, broad absorption component: a CBLA candidate system at $1314.2$ \AA.
From the simultaneous modeling of the COS and STIS data we find 
log $N$(H\,{\sc i}$)=13.27\pm0.13$ and $b=95\pm25$ km\,s$^{-1}$
as preferred solution for this CBLA candidate. This model is indicated with the 
red-shaded area in the lower right panels of Fig.\,7. Studying the
COS data alone, Danforth et al.\,(2016) also attributes the extended red
wing in the Ly\,$\alpha$ absorption to a BLA and derives
log $N$(H\,{\sc i}$)=13.95\pm0.15$ and $b=92\pm12$ km\,s$^{-1}$.
The STIS data analyzed here, however, favours a lower logarithmic column density 
than the 13.95 derived by Danforth et al.


\begin{deluxetable*}{lllrrrrr}
\tabletypesize{\scriptsize}
\tablewidth{0pt}
\tablecaption{Modeling results for CBLA candidates}
\tablehead{
\colhead{QSO name} &
\colhead{galaxy name} &
\colhead{$(L/L^{\star})_{\rm gal}$} &
\colhead{$cz_{\rm gal}$} &
\colhead{$cz_{\rm CBLA}$} &
\colhead{$D/R_V$} &
\colhead{log $N$(H\,{\sc i}$)_{\rm CBLA}$} &
\colhead{$b$(H\,{\sc i}$)_{\rm CBLA}$} \\
\colhead{} & \colhead{} & \colhead{} & \colhead{[km\,s$^{-1}$]} &
\colhead{[km\,s$^{-1}$]} & \colhead{} & \colhead{} &
\colhead{[km\,s$^{-1}$]} \\
}
\startdata
RX\,J\,1230.8+0115 & 2dFGRS-TGN388Z087         & 0.56  & 28480 & 28293 & 0.77 & $13.14\pm 0.16$ &  $90\pm20$ \\
PHL\,1811          & 2MASS\,J21545996-0922249  & 2.90  & 24283 & 24294 & 0.15 & $13.27\pm 0.13$ &  $95\pm25$ \\
3C\,273            & SDSS\,J122950.57+020153.7 & 0.02  &  1499 &  1684 & 0.75 & $13.16\pm 0.19$ & $120\pm15$ \\
PG\,1116+215       & SDSS\,J111906.68+211828.7 & 2.90  & 41671 & 41591 & 0.55 & $13.30\pm 0.22$ & $150\pm20$
\enddata
\end{deluxetable*}


The prediction from our CBLA model is
log $N$(H\,{\sc i}$)=13.20$, $b=180$ km\,s$^{-1}$ for 
$L\approx 2.9 L^{\star}$ and $D=34$ kpc, thus substantially broader
at a comparable column density. While it is possible to force
a very broad CBLA with $b=180$ km\,s$^{-1}$ and log $N$(H\,{\sc i}$)=13.20$
in our component model with an acceptable match between model and observations,
it is not the preferred solution of our modeling analysis. Possibly, the 
observed broad H\,{\sc i} feature stems from a somewhat cooler region 
in the MASS\,J21545996-0922249 halo with sub-virial temperatures (see also
Sect.\,5.6).

An alternative interpretation is that this CBLA candidate belongs to 
the somewhat fainter (and less massive) companion galaxy 2MASS\,J21545870-0923061, 
which has an impact parameter of $D=87$ kpc to the PHL 1811\,sightline at a redshift 
that is nearly identical to the closer 2MASS\,J21545996-0922249 galaxy.

\subsection{The CBLA candidate at $z=0.005$ towards 3C\,273}

Towards the optically brightest QSO on the sky, 3C\,273 ($z_{\rm em}=0.158, V=14.83$),
a very nearby CBLA candidate is detected at $z_{\rm abs}=0.005$,
probably related to hot gas in the outer halo of the $0.02 L^{\star}$ galaxy 
SDSS\,J122950.57+020153.7 at $D=80$ kpc (Stocke et al.\,2013) or
its intergalactic environment. At this redshift, the 3C\,273
sightline passes the outskirts of the Virgo cluster in a region
of substantial galaxy overdensity, but lies beyond the 
X-ray emission contours.

Because of the brightness of 3C\,273, both the COS and
STIS data are of excellent quality (S/N is $98$ per resolution element 
in the COS data and $24$ in the STIS data near 1225 \AA). 
The main H\,{\sc i} absorption component at 
$1222.1$ \AA\, is accompanied by a very broad, shallow absorption feature
centered near $1222.5$ \AA\, (see Fig.\,8, upper left two panels). 
This feature, although quite prominent, has been interpreted as 
continuum undulation in previous studies, but not as potential
Ly\,$\alpha$ absorption feature. In the detailed analysis of the
STIS spectral data of 3C\,273 (Tripp et al.\,2002; Williger et al.\,2010),
the very broad, shallow red wing of the strong Ly\,$\alpha$ absorption at
$z=0.00530$, that is visible in the raw STIS data (Tripp et al.\,2002; their 
Fig.\,1), is fitted as part of the continuum. In the superb COS data, this 
shallow feature is even more prominent (Fig.\,8) and distinct from a 
continuum undulation. Also other COS pipeline extractions for 3C\,273 
show this broad, shallow feature redwards of the main H\,{\sc i} absorption 
component, but strength and shape vary for these different data sets.
This feature is not considered in previous COS absorption-line studies 
(e.g., Danforth et al.\,2016).

We model this CBLA candidate based on the combined STIS/COS data set
of 3C\,273 with the parameters log $N$(H\,{\sc i}$)=13.16\pm0.19$ and 
$b=120\pm15$ km\,s$^{-1}$ (red-shaded area in Fig.\,8, upper right panels).
In our theoretical CBLA model grid we only consider galaxies with 
masses log $M\geq 10.5$ and luminosities $L\geq 0.1 L^{\star}$, so that 
we do not have any CBLA model prediction for this faint dwarf galaxy. 

The fact that such a low-luminosity galaxy ($R_{\rm vir}=106$ kpc) shows
such a prominent CBLA possibly points towards an extra-coronal origin
of the absorbing hot gas (assuming that the absorption feature is real). 
In view of the location of SDSS\,J122950.57+020153.7 within the Virgo environment, 
the CBLA may trace hot gaseous material that has been 
accumulated by the galaxy from the intracluster medium during the passage 
through the outer Virgo cluster. This example underlines the potential importance
of the interface regions between galaxy halos and their specific intergalactic 
environment, which may be studied best using constrained hydrodynamic cosmological
simulations of nearby galaxy filaments (e.g., Nuza et al.\,2015).

\subsection{The CBLA candidate at $z=0.139$ towards PG\,1116+215}

The sightline towards the Seyfert 1 galaxy PG\,1116+215
($z_{\rm em}=0.176, V=14.80$) passes the $2.90 L^{\star}$ galaxy
SDSS\,J111906.68+211828.7 ($z=0.139$) at an impact parameter of 
$D=138$ kpc (Stocke et al.\,2013).

A prominent, strong Ly\,$\alpha$ absorber is seen at the same
redshift as the galaxy together with various low, intermediate,
and high metal ions (e.g., Savage et al.\,2014; their Fig.\,10).
This system obviously traces a complex, multi-component and multi-phase
CGM absorber in the halo of SDSS\,J111906.68+211828.7. This is
an example that is representative for the difficulty of identifying 
broad absorption components in multi-phase absorbers. 
Previous studies of the COS data of this system has lead to some controversial 
results. Bluewards of the main Ly\,$\alpha$ absorption component at
$1384$ \AA\, there is a flux depression in the COS data 
(S/N is $50$ per resolution element at 1385 \AA)
that extends to $1383$ \AA. This feature is clearly present in
different pipeline extractions of the same COS data, although
different noise characteristics are evident (e.g., compare
Savage et al.\,2014, their Fig.\,10, with Stocke et al.\,2014, their
Fig.\,13). This flux depression is not considered in the multi-component
fit presented in Savage et al.\,(2014), but is fitted as a BLA with
$b=86\pm11$ km\,s$^{-1}$ down to $1383.3$ \AA\,in Stocke et al.\,(2014),
potentially being aligned with the strong (and broad) O\,{\sc vi}
absorption.

Interestingly, also the STIS data, that has a S/N of $18$ per resolution 
element at 1385 \AA, provides independent evidence for an extended wing
bluewards of the main Ly\,$\alpha$ component (see Fig.\,8). Combining
the COS and STIS data sets and considering the full extent of the
flux depression down to $1383.0$ \AA, we derive for this well-hidden 
CBLA candidate log $N$(H\,{\sc i}$)=13.30\pm0.22$ and
$b=150\pm20$ km\,s$^{-1}$ (red-shaded area in the lower right panel
of Fig.\,8). This solution also provides a better fit of the {\it red} wing of
the Ly\,$\alpha$ absorption, where a kink is seen in the COS
data near $1384.5$ \AA. While we cannot claim that this fit
is the definite solution for the decomposition of this particularly
complex CGM absorber, the combination of the STIS and COS data clearly
suggests that there is convincing evidence for a CBLA in this system 
that could be related to the coronal gas of SDSS\,J111906.68+211828.7.

The CBLA-model prediction is log $N$(H\,{\sc i}$)=13.06$ and 
$b=138$ km\,s$^{-1}$ for $L\approx 2.9 L^{\star}$ and $D=138$ kpc,
thus supporting the above given interpretation of the COS/STIS data.

\subsection{Interpretation of observed trends}

In all four examples presented above, the inclusion of a broad, shallow absorption component 
in the absorber models is required to account for the observed flux depressions in
the wings of the strong Ly\,$\alpha$ absorption and to provide an optimum fit to the COS and 
STIS spectral data. Also other authors have identified and discussed broad Ly\,$\alpha$ 
absorption as potential tracers for warm/hot circumgalactic gas in their analyses 
(e.g., Narayanan, Savage, Wakker 2010, 2012; Savage et al.\,2011; Savage et al.\,2014; 
Stocke et al.\,2014; Johnson et al.\,2017), but a systematic investigation of these
features with regard to the {\it expected} spectral signatures of shock-heated 
coronal gas has not been provided so far. Our study suggests that some of these
previously identified broad features may be indeed related to hot coronal gas around galaxies
and our CBLA models provide the theoretical basis for such a systematic study.

Within the errors from the CBLA modeling and the spectral analysis, the CBLA 
column densities and $b$ values derived from the COS/STIS data for the above-presented four 
CBLA candidate systems roughly agree with the CBLA model predictions. However, 
in view of the many additional systematic uncertainties involved (data-reduction issues, 
continuum undulations, lack of spectral resolution, limited S/N, etc.), a much larger 
sample of CBLA candidate systems is required to assess the relation between
hot coronal gas and broad H\,{\sc i} Ly\,$\alpha$ absorption in a statistically relevant 
manner.

One preliminary trend that we see in our four CBLA candidate systems is that 
the fitted H\,{\sc i} column densities are systematically higher ($0.2-0.5$ dex) 
than what is predictied by our CBLA models.
This discrepancy, if real, may be related to H\,{\sc i}-absorbing gas that resides in 
other (cooler) CGM phases or in warm-hot gas residing in the IGM beyond the virial radii. 
These and other aspects will be further investigated by us in our follow-up CBLA survey
and in future high-resolution CGM/IGM simulations, from which we will extract synthetic CBLA 
spectra.

%
\section{Summary and conclusions}

In this study, we have demonstrated that hot coronal gas in the extended halos of 
(predominantly) massive galaxies are expected to give rise to a weak but detectable Ly\,$\alpha$ 
absorption signal in the spectra of background AGN. The resulting absorber 
population, the CBLAs, may be used study the hot phase of the CGM in individual 
galaxies and/or to explore mass and extent of hot gas around galaxies in a 
statistical manner.

Our semi-analytic model predicts that CBLAs at $z\approx0$ span a characteristic 
H\,{\sc i} column-density/Doppler-parameter range of log $N$(H\,{\sc i}$)=12.4-13.4$ 
and $b=70-200$ km\,s$^{-1}$. As we have demonstrated, such broad, shallow 
absorption features at low redshift are detectable only in high-S/N UV spectral data,
but even there they may be hidden within the overall (often complex) Ly\,$\alpha$ 
absorption pattern that is usually dominated by cooler CGM gas components.
A careful modeling of the Ly\,$\alpha$ absorption profiles is required to identify 
CBLA candidate systems in CGM absorbers.

We have provided four examples for such a CBLA modeling by combining 
archival HST/COS and HST/STIS data. 
The inclusion of a CBLA component in the spectral models
is required to provide a satisfying fit to the COS/STIS data,
such as it is also seen in CGM systems studied by other
authors (Narayanan, Savage, Wakker 2010, 2012; Savage et al.\,2011, 2014;
Stocke et al.\,2014; Johnson et al.\,2017).
Although blending effects and the 
limited S/N complicate the interpretation of the observed features in our four 
example spectra, the modeled line profiles qualitatively match the expected 
CBLA characteristics.
There appears to be a mild ($0.2-0.5$ dex) excess in H\,{\sc i} column density 
seen in the COS/STIS data when compared to the CBLA model predictions. 
This could be related to H\,{\sc i}-absorbing gas residing in other CGM phases 
or in the IGM outside the galaxies' virial radii. A larger CBLA candidate
sample will be required to further investigate these aspects in more detail.

One important conclusion from our study is that profile-fitting of CGM absorbers 
{\it generally} should take into account the possible presence of a CBLA
absorption component. With this study, we
provide a parametrization of the expected column densities and $b$ values
of CBLAs as a function of halo mass and impact parameter. This might be 
useful to model the expected shape of CBLAs in absorption systems in future 
CGM studies.

Given the fact that the expected number density per unit redshift of CBLAs at $z\approx0$ 
is relatively large ($d{\cal N}/dz\approx 3$), a survey 
of CBLAs in galaxy-halo sightlines at $z\approx 0$ sampled with HST/COS is the
next logical step to further explore the nature of these systems. Such a CBLA
survey will be presented by us in a future paper together with a careful assessment
of systematic uncertainties in the HST/COS data (fixed-pattern noise, instrumental
artifacts, continuum undulations, etc.).
  
Another important future project will be the systematic exploration of broad H\,{\sc i} 
featueres arising in the $T=10^6$ K phase in synthetic spectra from numerical
hydrodynamic CGM simulations of galaxies with different masses and 
evolutionary states. As mentioned earlier, our idealized, analytic halo model 
assumes a specific temperature/density profile that does not account for feedback processes 
or other crucial aspects of galaxy formation/evolution.
The relevance of these aspects for the occurance and shape of broad H\,{\sc i} 
Ly\,$\alpha$ features in galaxy halos can only can be studied based 
on state-of-the-art numerical simulations that cover the cosmological framework, 
the necessary gas physics, and the required high spatial resolution.
Note that previous CGM simulations have already indicated the presence of broad H\,{\sc i}
lines that represent the analogs of CBLAs (e.g., Liang, Kravtsov \& Agertz 2018; their Fig.\,3).

High-resolution CGM simulations will further provide crucial information on 
potential temperature fluctuations in the inner and
outer corona and the role of (non-thermal) bulk motions in the coronal gas (e.g., from outflows
and mergers) for the H\,{\sc i} line-broadening. They also will be essential to 
characterize the transition zone between the CGM of individual galaxies and the IGM in 
the superordinate cosmologial environment (filaments, galaxy groups) that contains
shock-heated hot gas as well (see, e.g., Stocke et al.\,2014; Nuza et al.\,2015;
Bouma, Richter \& Fechner 2019).

%

\acknowledgements

Most of the research presented in this paper has been carried out at the
Department of Physics and Astronomy of the University of Canterbury,
Christchurch, New Zeland, during may stay as guest professor and
visiting Erskine fellow between Feb and June 2019.
I am extremely grateful for the financial and organisational support of the 
University of Canterbury in the framework of this fellowship.
I would also like to thank Andy Fox and Nicolas Lehner for helpful comments
and remarks.

%

\input{references2.tex}

%

\newpage

\appendix

In this Appendix, we present the most relevant equations for the semi-analytic 
CBLA modeling presented in Sect.\,3.\\

\section{The coronal gas distribution in individual galaxy halos}

To characterize the spatial extent, radial density distribution, and radial temperature 
distribution of the hot coronal gas in a DM halo of given virial
mass, $M_{\rm V}$, we use the formalism outlined in the seminal paper presented
by Maller \& Bullock in 2004 (MB04).

\subsection{DM halo properties}

Following the approach presented in MB04, we calculate the virial radius, $R_{\rm V}$, 
for a galaxy with viral mass, $M_{\rm V}$, via the relation

\begin{equation}
R_{\rm V}=206\,h^{-1}\,{\rm kpc}\,\left(
\frac{\Delta_{\rm V}\Omega_{\rm m}}{97.2}\right)^{-1/3}\,
\left(\frac{M_{\rm V}}{10^{12}\,h^{-1}\,M_{\sun}}\right)^{1/3}\,(z+1)^{-1},
\end{equation}

where $\Omega_{\rm m}$ is the cosmological matter-density parameter and $\Delta_{\rm V}$
is the virial overdensity, here set to $\Delta_{\rm V}\equiv 200$. 

Cosmological $N$-body simulations have demonstrated that the matter density
in a DM halo follows a radial profile function in the form

\begin{equation}
\rho(R)=\frac{\rho_{\rm S}\,R_{\rm S}^{\,3}}{R(R+R_{\rm S})^2}.
\end{equation}

This is the Navarro-Frenk-White (NFW) profile (Navarro, Frenk \& White 1995;
Klypin et al.\,2001) with the parameters $\rho_{\rm S}$ as characteristic
density and $R_{\rm S}$ as scale radius. Scale radius and virial radius 
in a NFW DM halo are connected via the so-called concentration parameter,
$C_{\rm V}=R_{\rm V}/R_{\rm S}$, where $C_{\rm V}$ can be approximated
via the relation $C_{\rm V}=9.6\,(M_{\rm V}/10^{13}M_{\sun})^{-0.13}\,(1+z)^{-1}$
(Bullock et al.\,2001).
For the maximum circular velocity in a NFW profile we can write

\begin{equation}
V_{\rm max}=\sqrt{\frac{GM(R_{\rm max})}{R_{\rm max}}},
\end{equation}

where $R_{\rm max}\approx 2.15R_{\rm S}$.

\subsection{Initial gas density profile and coronal gas temperature}

We now consider the distribution of hot (virialized) gas confined in a NFW DM potential well. 
We assume that the baryonic mass fraction in the seed halo (initially purely in the
form of gas) is tied to the cosmological baryon fraction, $f_{\rm b}$, so
that $M_{\rm b}=f_{\rm b}M_{\rm V}$. Following MB04, we can write for the 
radial mass-density profile of the hot gas

\begin{equation}
\rho_{\rm cor}(R)=\frac{R_{\rm S}^{\,3}\,\rho_0}
{[R+0.75R_{\rm S}](R+R_{\rm S})^2}.
\end{equation}

The core density, $\rho_0$, depends on the total baryonic gas mass and the 
concentration parameter in the way

\begin{equation}
\rho_0=\frac{M_{\rm b}}{4\pi R_{\rm S}^{\,3}\,g(C_{\rm V})},
\end{equation}

where the function $g(x)$ has the form

\begin{equation}
g(x)=9\,{\rm ln}\,\left(1+\frac{4}{3}x\right)-8\,{\rm ln}\,(1+x)
-\frac{4x}{1+x}.
\end{equation}

The initial temperature of the hot, coronal (isothermal) gas, $T_{\rm cor}$, depends
on the sound speed, $c_{\rm s}=V_{\rm max}/\sqrt{2}$, and can be written as

\begin{equation}
T_{\rm cor}=\frac{\mu_{\rm i}m_{\rm p}c_{\rm s}^2}{\gamma k_{\rm B}}.
\end{equation}

Here, $\mu_{\rm i}=0.62$ is the mean mass per particle in the fully 
ionized plasma (assuming a helium mass fraction of 30 percent),
$m_{\rm p}$ is the proton mass, $\gamma$ is the polytropic index
(assumed to be unity for an ideal isothermal gas), and $k_{\rm B}$
is the Boltzmann constant.

\subsection{Gas cooling and the  multi-phase nature of the CGM}

In the inner region of the halo, where the gas density is the highest, the
hot gas is able to cool within a Hubble time. 
With $\mu_{\rm e}=1.18$ as the mean mass per electron, and $\Lambda(T,Z_{\rm g})$
as cooling function, the density threshold above which the coronal gas is able to
cool in the time scale $t_{\rm f}$ is given by

\begin{equation}
\rho_{\rm c}=\frac{3\mu_{\rm e}^2m_{\rm p}k_{\rm b}T}
{2\mu_{\rm i}t_{\rm f}\Lambda(T,Z_{\rm g})}.
\end{equation}

The time scale that is relevant here is the {\it halo-formation time scale},
assumed to be $t_{\rm f}=8$ Gyr (see discussion in MB04).
The cooling function, $\Lambda$, which depends mainly on the temperature, $T$,
and the overall metallicity of the gas, can be approximated by a simple power law:

\begin{equation}
\Lambda(T,Z_{\rm g})=2.6\times 10^{-23}\,\Lambda_{\rm Z}\,
\left(\frac{T}{10^6\,{\rm K}}\right)^{-1}\,{\rm cm}^3{\rm erg\,s}^{-1},
\end{equation}

where the cooling parameter $\Lambda_{\rm Z}$ scales with with the gas
metallicity, $Z_{\rm g}$ (in solar units), as given in MB04 (their Table A.1).
We here generally assume $Z_{\rm g}=0.1$.

The density threshold for cooling gas in a DM halo corresponds to a 
characteristic radius, often referred to as the {\it cooling radius}, $R_{\rm c}$
(Frenk \& White 1991). According to MB04, this radius can approximated by the
relation

\begin{equation}
R_{\rm c}\approx 157\,{\rm kpc}\,\left(\frac{T}{10^6\,{\rm K}}\right)^{-1/8}\,
\left(\frac{\Lambda_{\rm Z}\,t_{\rm f}}{8\times 10^9\,{\rm yr}}\right)^{1/3}.
\end{equation}

At $R\leq R_{\rm c}$, the hot coronal gas will cool and fragment, leading to
star formation in the inner-most region of the galaxy and a multi-phase, inner CGM,
whereas for $R>R_{\rm c}$ the gas is assumed to remain hot and isothermal at 
$T=T_{\rm cor}$. Also at $R\leq R_{\rm c}$, a certain fraction of the CGM will 
be hot, however, as the gas is multi-phase. MB04 derive expressions for the 
density and temperature profiles of this residual hot gas component at
$R\leq R_{\rm c}$ under the assumption, that the gas reaches hydrostatic
equilibrium and responds adiabatically to pressure changes:

\begin{equation}
\rho_{\rm res}(R\leq R_{\rm c})=\rho_{\rm cor}\,\left[
1+\frac{3.7R_{\rm S}}{R}\,{\rm ln}\left(1+\frac{R}{R_{\rm S}}\right)-
\frac{3.7R_{\rm S}}{R_{\rm c}}\,{\rm ln}\left(1+\frac{R_{\rm c}}{R_{\rm S}}
\right)\right]^{3/2}
\end{equation}

and

\begin{equation}
T_{\rm res}(R\leq R_{\rm c})=T_{\rm cor}\,\left[
1+\frac{3.7R_{\rm S}}{R}\,{\rm ln}\left(1+\frac{R}{R_{\rm S}}\right)-
\frac{3.7R_{\rm S}}{R_{\rm c}}\,{\rm ln}\left(1+\frac{R_{\rm c}}{R_{\rm S}}
\right)\right].
\end{equation}

With equations (A4), (A7) for $R>R_{\rm c}$ and (A11), (A12) for $R\leq R_{\rm c}$
we are now able to calculate for each galaxy halo with virial mass $M_{\rm V}$
the radial gas density and temperature distributions, $\rho(R)$ and $T(R)$, of the
hot coronal gas. Finally, we convert the mass density $\rho(R)$ into an hydrogen
particle density using the relation $n_{\rm H}(R)=\rho(R)/\mu_{\rm i}m_{\rm p}$. 

\subsection{Ly\,$\alpha$ absorption properties}

For a weak, unsaturated absorption line, there is a simple, linear relation between
the observed equivalent width and the absorbing gas column density:

\begin{equation}
\left(\frac{W_{\lambda}}{\rm \AA}\right)=8.85\times 10^{-21}\,f\,\left(\frac{N}{{\rm cm}^{-2}}\right)
\left(\frac{\lambda_0}{\rm \AA}\right)^2.
\end{equation}

Here, $\lambda_0$ denotes the laboratory wavelength of the transition and
$f$ its oscillator strength. With $\lambda_0=1215.67$ \AA\,and $f=0.4164$
for the H\,{\sc i} Ly\,$\alpha$ line (Morton 2003) we get:

\begin{equation}
\left[ \frac{W_{\lambda}({\rm HI, Ly}\alpha)}{\rm \AA} \right] = 
0.545\,\left[\frac{N({\rm HI, Ly}\alpha)}{10^{14}\,{\rm cm}^2}\right].
\end{equation}

We now calculate the central absorption depth, $\cal{D}$, of the CBLAs as a function 
of $N$(H\,{\sc i}) and $b$ value. For a Gaussian-shaped absorption line,
there is a simple relation between the equivalent width (i.e., the area
under the Gaussian profile), the full-width-at-half-maximum ($\Delta_{\rm FWHM}$) of the 
Gaussian, and the central absorption depth:

\begin{equation}
W_{\lambda}=\frac{\sqrt{2\pi}\,\Delta_{\rm FWHM}\,\cal{D}}{2\sqrt{2\,{\rm ln}\,2}}.
\end{equation}

Taking into account that for a Gaussian-shaped line $\Delta_{\rm FWHM}=1.66\,b$ and
plugging in the correct numbers for H\,{\sc i} Ly\,$\alpha$ to convert between
wavelength and velocity space, we obtain

\begin{equation}
{\cal D}=139\,\left[ \frac{W_{\lambda}({\rm HI, Ly}\alpha)}{\rm \AA} \right]\,
\left[\frac{b({\rm HI})}{{\rm km\,s}^{-1}}\right]^{-1}. 
\end{equation}


\begin{figure}[t!]
\epsscale{0.40}
\plotone{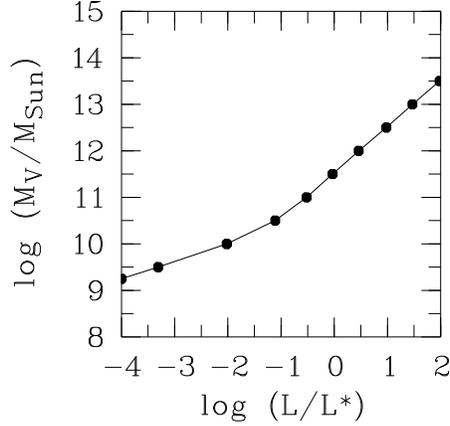}
\caption{Assumed mass-luminosity relation for $z=0$ galaxies, based on
the studies of Stocke et al.\,(2014) and Moster et al.\,(2010).
}
\end{figure}


\section{The cosmological cross section of coronal gas in the local Universe}

We now discuss the equations that we need to determine the 
cosmological cross section of coronal gas halos and their CBLA signatures 
at low redshift. For this, we make use of the local galaxy luminosity function and
standard cosmological equations.

\subsection{Space density of galaxy halos}

The space density of galaxies per unit luminosity is given by the Schechter
luminosty function (Schechter 1976), which has the form

\begin{equation}
\phi(L)=\left(\frac{\phi^{\star}}{L^{\star}}\right)\,\left(\frac{L}{L^{\star}}\right)^{\alpha}
\,e^{-(L/L^{\star})}.
\end{equation}

Here, $L^{\star}$ is a characteristic luminosity, $\alpha$ is the slope at the faint end of the
luminosity function, and $\phi^{\star}$ is the normalization density.
The space density of galaxies with luminosities $L'\geq L$ is given by the integral

\begin{equation}
\phi(L'>L)=\int_{L}^{\infty}\,\phi(L)\,{\rm d}L=
\phi^{\star}\,\int_{(L/L^{\star})}^{\infty}\left(\frac{L}{L^{\star}}\right)^{\alpha}
\,e^{-(L/L^{\star})}\,{\rm d}\left(\frac{L}{L^{\star}}\right)=
\phi^{\star}\,\Gamma(\alpha+1,L/L^{\star}),
\end{equation}

where $\Gamma$ stands for the upper incomplete Gamma function with the arguments
$\alpha+1$ and $L/L^{\star}$.

By solving equation (B2) numerically via the Gamma function, we obtain the space density
of galaxies per luminosity bin, $\phi(\Delta L)$, in units [$h^{-3}$Mpc$^{-3}$]. The
parameters $\phi^{\star}$, $\alpha$, and $L^{\star}$ are obtained from observations of
the local galaxy luminosity function (Montero-Dorta \& Prada 2009).
To convert this into a space density as a function of galaxy's virial halo mass, we
use the mass-luminosity relation presented in Stocke et al.\,(2014), which is
based on halo-matching models of Moster et al.\,(2010). The relation between
mass and luminosity (Fig.\,A1) can be approximated by the polynomial

\begin{equation}
{\rm log}\,\left(\frac{M_{\rm V}}{M_{\sun}}\right)=
11.500+0.991\,X+0.085\,X^2-0.029\,X^3-0.006\,X^4,
\end{equation}

where $X={\rm log}\,(L/L^{\star})$.
The combination of equations (B2) and (B3) allows us to derive the space density of galaxies
and their halos in a given (virial) mass bin, $\phi(\Delta M_{\rm V})$, which is used to 
calculate the cross section of coronal gas for the local galaxy population (Sect.\,4.4).


\section{Isothermal model}

In Fig.\,C.1, we show the dependence of CBLA H\,{\sc i} column densities and $b$ values on
the virial halo mass for an {\it isothermal} 
halo model, where we set $T=T_{\rm cor}(M_V)$ for each halo mass.
Compared to the MB04 model (Fig.\,4), the isothermal model gives sytematically
higher H\,{\sc i} column densities owing to the higher neutral gas fractions
(Eq. 3) in the now somewhat cooler inner halo regions. 

In contrast to the MB04 model (Fig.\,4), the Doppler parameter $b$(CBLA) in the 
isothermal model is constant for each halo mass and directly related to the virial halo 
temperature via Eq. (6). The resulting absorption lines thus would be substantally 
narrower and easier to detect (Sect.\,4.2).

Fig.\,4 and Fig.\,C.1 can be compared to H\,{\sc i} column density and $b$ value
distributions from CBLAs seen in synthethic UV spectra generated from high-resolution
CGM simulations (e.g., Liang, Kravtsov \& Agertz 2018).


\begin{figure}[t!]
\epsscale{0.40}
\plotone{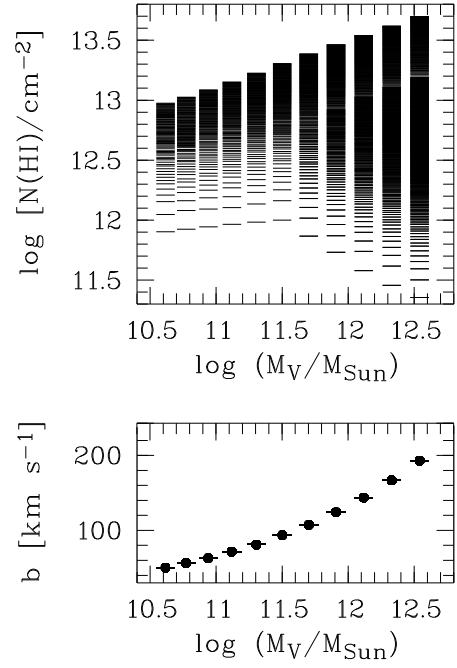}
\caption{Dependence of CBLA H\,{\sc i} column densities and $b$ values on
the virial halo mass for an isothermal halo model (compare to Fig.\,4).
}
\end{figure}

%

\end{document}

%% file: references2.tex
\section*{REFERENCES}
\begin{footnotesize}

\noindent
\\
Anderson, M.E., \& Bregman, J.N. 2010, ApJ, 714, 320
\noindent
\\
Anderson, M.E., \& Bregman, J.N. 2011, ApJ, 737, 22
\noindent
\\
Anderson M.E., Churazov E., Bregman J.N., 2016, MNRAS, 455, 227
\noindent
\\
Borthakur, S., Heckman, T., Tumlinson, J., et al.\,2016, ApJ, 833, 259
\noindent
\\
Bouma, S.J.D., Richter, P., Fechner, C. 2019, A\&A, 627, A20
\noindent
\\
Bregman, J.N., \& Houck, J.C. 1997, ApJ, 485, 159
\noindent
\\
Bullock, J.S., Kolatt, T.S., Sigad, Y., Somerville, R.S., Kravtsov, A.V., 
Klypin, A.A., Primack, J.R., Dekel, A. 2001, MNRAS, 321, 559
\noindent
\\
Burchett, J.N., Tripp, T.M., Prochaska, J.X., et al.\,2019, ApJL, 877, 20
\noindent
\\
Danforth, C.W., Tilton, E.M., Shull, J.M., et al.\,2016, ApJ, 817, 111
\noindent
\\
Fang, T., McKee, C.F., Canizares, C.R., \& Wolfire, M. 2006, ApJ, 644, 174
\noindent
\\
Fontana, A., \& Ballester, P. 1995, ESO Messenger, 80, 37
\noindent
\\
Fukugita, M. \& Peebles, P.J.E. 2006, ApJ, 639, 590
\noindent
\\
Gutcke, T.A., Stinson, G.S., Macci\'o, A.V., Wang, L., \& Dutton, A.A. 2017, MNRAS, 464, 2796
\noindent
\\
Hani, M.H., Ellison, S.L., Sparre, M., Grand, R.J.J., Pakmor, R., Gomez, F.A., 
\& Springel, V. 2019, MNRAS, 488, 135
\noindent
\\
Haynes, M.P., Giovanelli, R., Martin, A.M., et al.\,2011, AJ, 142, 170
\noindent
\\
Hodges-Kluck, E.J., Miller, M.J., \& Bregman, J.N. 2016, ApJ, 822, 21
\noindent
\\
Jenkins, E.B., Bowen, D.V., Tripp, T.M., et al. 2003, AJ, 125, 2824
\noindent
\\
Jenkins, E.B., Bowen, D.V., Tripp, T.M., \& Sembach, K.R. 2005, ApJ, 623, 767
\noindent
\\
Johnson, S.D., Chen, W.-W., Mulchaey, J.S., Schaye, J. \& Straka, L.A. 2017, ApJ 850, L10
\noindent
\\
Keeney, B.A., Stocke, J.T., Pratt, C.T., et al.\,2018, ApJS, 237, 11
\noindent
\\
Klypin A., Kravtsov A.V., Bullock J.S., \& Primack J.R., 2001, ApJ, 554, 903
\noindent
\\
Li, J.-T., Li, Z., Wang, Q.D., Irwin, J.A., \& Rossa, J. 2008, MNRAS, 390, 59
\noindent
\\
Li J.-T., Bregman J.N., Wang Q.D., Crain R.A., \& Anderson M.E. 2016, ApJ, 830, 134
\noindent
\\
Li J.-T. \& Bregman J.N. 2017, ApJ, 849, L105
\noindent
\\
Lehner, N., Savage, B. D., Richer, P., et al. 2007, ApJ, 658, 680
\noindent
\\
Liang, C.J., \& Chen, H.-W. 2014, MNRAS, 445, 2061
\noindent
\\
Liang, C.J., Kravtsov, A.V., \& Agertz, O. 2018, MNRAS, 479, 1822
\noindent
\\
Maller, A.H., \& Bullock, J.S. 2004, MNRAS, 355, 694 (MB04)
\noindent
\\
McCammon, D., Almy, R., Apodaca, E., et al.\,2002, ApJ, 576, 188
\noindent
\\
Miller, M.J. \& Bregman, J.N. 2013, ApJ, 770, 118
\noindent
\\
Miller, M.J. \& Bregman, J.N. 2015, ApJ, 800, 14
\noindent
\\
Morton, D.C. 2003, ApJS, 149, 205
\noindent
\\
Montero-Dorta, A.D., \& Prada, F. 2009, 399, 1106
\noindent
\\
Moster, B.P., Somerville, R.S., Maulbetsch, C., van\,den\,Bosch, F.C.,
Maccio, A.V., Naab, T., \& Oser, L. 2010, ApJ, 710, 903
\noindent
\\
Muzahid, S., Fonseca, G., Roberts, A., Rosenwasser, B., Richter, P.,
Narayanan, A., Churchill, C., Charlton, J. 2018, MNRAS, 476, 4965
\noindent
\\
Navarro J.F., Frenk C.S., \& White S.D.M. 1995, MNRAS, 275, 56
\noindent
\\
Narayanan, A., Savage, B.D., \& Wakker, B.P. 2012, ApJ, 752, 65
\noindent
\\
Nicastro, F., Zezas, A., Drake, J., et al.\,2002, ApJ, 573, 157
\noindent
\\
Nuza, S.E., Parisi, F., \& Scannapieco, C., et al.\,2014, MNRAS, 441, 2593
\noindent
\\
O’Sullivan, E., Ponman, T.J., \& Collins, R.S. 2003, MNRAS, 340, 1375
\noindent
\\
Paerels, F.B.S. \& Kahn, S.M. 2003, ARA\&A, 41, 291
\noindent
\\
Pisano, D.J., Barnes, D.G., Gibson, B.K., Staveley-Smith, L., Freeman, K.C., 
\& Kilborn, V.A. 2007, ApJ, 662, 959
\noindent
\\
Prause, N., Reimers, D., Fechner, C., \& Janknecht, E., 2007, A\&A, 470, 67
\noindent
\\
Prochaska, J.X., Weiner, B., Chen, H.-W., Mulchaey, J., \& Cooksey, K. 2011, ApJ, 740, 91
\noindent
\\
Prochaska, J.X., Burchett, J.N., Tripp, T.M., et al.\,2019, ApJ, 243, 21
\noindent
\\
Rasmussen, J., Sommer-Larsen, J., Pedersen, K., Toft, S., 
Benson, A., Bower, R.G., \& Grove, L.F. 2009, ApJ, 697, 79
\noindent
\\
Richter, P., Savage, B.D., Tripp, T.M., \& Sembach, K.R. 2004, ApJS, 153, 165
\noindent
\\
Richter, P., Savage, B.D., Sembach, K.R., \& Tripp, T.M. 2006, A\&A, 445, 827
\noindent
\\
Richter, P., Fang, T., \& Bryan, G. L. 2006, A\&A, 451, 767
\noindent
\\
Richter, P., Paerels, F.B.S., Kaastra, J.S. 2008, SSRv, 134, 25
\noindent
\\
Richter, P., Charlton, J.C., Fangano, A.P.M., Ben Bekhti, N., \& Masiero, J.R. 2009, ApJ, 695, 1631
\noindent
\\
Richter, P. 2012, ApJ, 750, 165
\noindent
\\
Richter, P., Fox, A.J., Wakker, B.P., Lehner, N., Howk, J.C., Bland-Hawthorn, J., 
Ben\,Bekhti, N., \& Fechner, C. 2013, ApJ, 772, 111
\noindent
\\
Richter, P., Fox, A.J., Ben Bekhti, N., Murphy, M.T., Bomans, D., \& Frank, S. 2014, AN, 335, 92
\noindent
\\
Richter, P., Wakker, B.P., Fechner, C., et al.\,2016, A\&A, 590, A68
\noindent
\\
Richter, P.; Nuza, S.E.; Fox, A.J. et al. 2017, A\&A, 607, A48
\noindent
\\
Richter, P. 2017, in: Gas Accretion onto Galaxies, Astrophysics and Space Science Library,
eds. A. J. Fox \& R. Dav{\'e} (Springer), 15
\noindent
\\
Richter, P.; Winkel, B.; Wakker, B.P. et al.\,2018, ApJ, 868, 112
\noindent
\\
Savage, B.D., Narayanan, A., Lehner, N. \& Wakker, B.P. 2012, ApJ, 731, 14
\noindent
\\
Savage, B.D., Kim, T.-S., Wakker, B.P., Keeney, B., Shull, J.M., Stocke, J.T., 
\& Green, J.C. 2014, ApJS, 212, 8
\noindent
\\
Sembach, K. R., Tripp, T. M., Savage, B. D., \& Richter, P. 2004, ApJS, 155, 351
\noindent
\\
Schechter, P. 1976, ApJ, 203, 297
\noindent
\\
Spitzer, L. 1956, ApJ, 124, 20
\noindent
\\
Stocke, J.T., Keeney, B. A., Danforth, C. W., et al. 2013, ApJ, 763, 148
\noindent
\\
Stocke, J.T., Keeney, B.A., Danforth, C.W., et al.\,2014, ApJ, 791, 128
\noindent
\\
Strickland, D.K., Heckman, T.M., Colbert, E.J.M., Hoopes, C.G., \& Weaver, K.A. 2004, ApJS, 151, 193
\noindent
\\
Tepper-Garc{\'i}a, T., Richter, P., Schaye, J., et al.\,2012, MNRAS, 425, 1640
\noindent
\\
Tripp et al. 2009
\noindent
\\
T\"ullmann, R., Pietsch, W., Rossa, J., Breitschwerdt, D., \& Dettmar, R.-J. 2006, A\&A, 448, 43
\noindent
\\
Tumlinson, J., Thom, C., Werk, J.K., et al.\,2013, ApJ, 777, 59
\noindent
\\
Tumlinson, J., Peeples, M.S., Werk, J.K. 2017, ARA\&A, 55, 389
\noindent
\\
van\,de\,Voort, F., Springel, V., Mandelker, N., van\,den\,Bosch, F.C., \& Pakmor, R. 2019, MNRAS, 482, L85
\noindent
\\
Wang, Q.D., Yao, Y., Tripp, T.M., et al.\,2005, ApJ, 635, 386
\noindent
\\
Wakker, B.P., \& Savage, B.D. 2009, ApJS, 182, 378
\noindent
\\
Werk, J.K., Prochaska, J.X., Thom, C., et al.\,2013, ApJS, 204, 17
\noindent
\\
White, S.D.M. \& Frenk, C.S. 1991, ApJ, 379, 52
\noindent
\\
Williams, R.J., Mathur, S., Nicastro, F., et al.\,2005, ApJ, 631, 856
\noindent
\\
York, D.C. \& Cowie, L.L. 1983, ApJ, 264, 49
\noindent
\\
Yun, M.S., Ho, P.T.P., \& Lo, K.Y. 1994, Nature, 372, 530

\end{footnotesize}